\documentclass[twocolumn,secnumarabic,amssymb, nobibnotes, aps, prl, superscriptaddress, nobalancelastpage,longbibliography]{revtex4-2}

 \usepackage{graphicx}
\usepackage{booktabs} 
\usepackage{multirow}
\usepackage{bm}
\usepackage{amsmath} 
\usepackage{braket} 
\usepackage{epsfig}
\usepackage{tensor}
\usepackage{CJKutf8}
\usepackage[version=4]{mhchem}
\usepackage{titlesec}
\usepackage{ifthen}
\usepackage{sidecap}
\usepackage{listings}
\usepackage{array}
\usepackage{floatrow}
\usepackage{enumitem}
\usepackage[para,online,flushleft]{threeparttablex}
\usepackage{diagbox}
\usepackage{float}
\floatstyle{plaintop}
\restylefloat{table}

\usepackage{hyperref}
\usepackage{xr}
\hypersetup{breaklinks=true,colorlinks=true,linkcolor=blue,citecolor=blue,filecolor=magenta,urlcolor=cyan}

\usepackage[all]{hypcap}

\usepackage{xcolor}
\definecolor{pastelgray}{rgb}{0.81, 0.81, 0.77}
\definecolor{beaublue}{rgb}{0.9, 0.9, 0.93}

\usepackage{orcidlink}
\usepackage{tikz,xcolor,hyperref}

\definecolor{lime}{HTML}{A6CE39}
\DeclareRobustCommand{\orcidicon}{
	\begin{tikzpicture}
	\draw[lime, fill=lime] (0,0) 
	circle [radius=0.16] 
	node[white] {{\fontfamily{qag}\selectfont \tiny ID}};
	\draw[white, fill=white] (-0.0625,0.095) 
	circle [radius=0.007];
	\end{tikzpicture}
	\hspace{-2mm}
}

\foreach \x in {A, ..., Z}{%
	\expandafter\xdef\csname orcid\x\endcsname{\noexpand\href{https://orcid.org/\csname orcidauthor\x\endcsname}{\noexpand\orcidicon}}
}

\newcolumntype{Y}{>{\centering\arraybackslash}X}

\begin{document}

\begin{CJK*}{UTF8}{gbsn}

\title{Three-body Effect in Short-range Correlations}




\author{H. Y. Shang\,\orcidlink{0009-0007-1253-4519}}
\affiliation{Key Laboratory of Nuclear Physics and Ion-beam Application (MOE), Institute of Modern Physics, Fudan University, Shanghai 200433, China}
\affiliation{Shanghai Research Center for Theoretical Nuclear Physics,
NSFC and Fudan University, Shanghai 200438, China}
\author{R. Z. Hu\,\orcidlink{0009-0002-8797-6622}}
\affiliation{
State Key Laboratory of Nuclear Physics and Technology, School of Physics,
Peking University, Beijing 100871, China
}
\author{X. Y. Xu\,\orcidlink{0009-0002-5053-0352}}
\affiliation{
State Key Laboratory of Nuclear Physics and Technology, School of Physics,
Peking University, Beijing 100871, China
}
\author{Z. C. Xu\,\orcidlink{0000-0001-5418-2717}}
\affiliation{Key Laboratory of Nuclear Physics and Ion-beam Application (MOE), Institute of Modern Physics, Fudan University, Shanghai 200433, China}
\affiliation{Shanghai Research Center for Theoretical Nuclear Physics,
NSFC and Fudan University, Shanghai 200438, China}
\author{J. C. Pei\,\orcidlink{0000-0002-9286-1304}}
\affiliation{
State Key Laboratory of Nuclear Physics and Technology, School of Physics,
Peking University, Beijing 100871, China
}
\affiliation{
Southern Center for Nuclear-Science Theory (SCNT), Institute of Modern Physics, Chinese Academy of Sciences, Huizhou 516000,  China
}
\author{S. M. Wang\,\orcidlink{0000-0002-8902-6842}}\email{Email: wangsimin@fudan.edu.cn}
\affiliation{Key Laboratory of Nuclear Physics and Ion-beam Application (MOE), Institute of Modern Physics, Fudan University, Shanghai 200433, China}
\affiliation{Shanghai Research Center for Theoretical Nuclear Physics,
NSFC and Fudan University, Shanghai 200438, China}

\date{\today}

\begin{abstract}
Short-range correlations (SRCs) provide the link between low- and high-energy nuclear physics and can be quantified by two-nucleon densities.
We present calculations of the two-nucleon densities using free-space similarity renormalization group (SRG)-evolved operators 
and in-medium SRG (IMSRG) ground states with softend chiral interaction. 
Our calculations benchmark well against no-core shell model (NCSM) results with unevolved oparetors and Hamiltonians in $^4\mathrm{He}$.
We explicitly include the induced three-body (3b) density operators for the first time which, together with the 3b Hamiltonians, provide the full 3b effects. 
We show pronounced 3b effects in the $^{16}\mathrm{O}$ two-nucleon densities. 
Combined with valence-space IMSRG (VS-IMSRG) method, we extend the calculation to the oxygen isotopic chain. 
This approach enables a consistent \textit{ab initio} description of low-energy properties and SRCs within one framework 
and offers predictions for the upcoming SRC measurements in unstable nuclei.
\end{abstract}

\maketitle


\emph{Introduction.}---
Nuclei are quantum many-body systems composed of nucleons, which themselves emerge from the color confinement of quarks. The structural properties of nuclei are governed by the residual strong interaction---the nuclear force---which acts between these composite nucleons. In this complex environment, nucleons are not merely independent particles moving in a mean field. Approximately 20\% of nucleons participate in short-range correlations (SRCs): transient, high-density fluctuations where two nucleons, predominantly a neutron-proton pair, approach within about 1\,fm of each other~\cite{hen2017nucleon}. These SRC pairs are governed by the strong, short-ranged tensor component of the nuclear interaction and carry large relative momenta~\cite{schmidt2020probing,korover2014probing,cai2025srceosdensematter}, dominating the high-momentum tails of nuclear wave functions. Their study provides a powerful bridge between low-energy nuclear structure and the underlying high-energy dynamics of quantum chromodynamics (QCD)~\cite{denniston2024modification}. Experiments using high-energy probes have established that SRCs not only reveal the nature of the nuclear interaction at short distances but are also phenomenologically linked to the modification of quark distributions in nuclei---the celebrated European Muon Collaboration (EMC) effect~\cite{weinstein2011short}. This connection suggests that the internal structure of nucleons is temporarily altered when they participate in these correlated, high-virtuality configurations.


Theoretical descriptions of SRCs rely on the two-nucleon density $\rho_2$~\cite{degli2015medium}, from which key observables like the SRC scaling factor $a_2$ and nuclear contacts are derived. The factor $a_2$, linearly related to the EMC effect~\cite{weinstein2011short}, quantifies the overall SRC abundance~\cite{chen2017short}, while contacts characterize the isospin structure of SRC pairs~\cite{weiss2015generalized,cruz2021many}. Extracting the high-momentum content of $\rho_2$ within a fully microscopic framework requires either a hard nuclear interaction~\cite{tropiano2021short,tropiano2024high,shang2025many} or a consistent renormalization-group evolution~\cite{bogner2007similarity,bogner2010low} of both the Hamiltonian and the density operator~\cite{stetcu2006long,anderson2010operator,neff2015short}. However, calculations with hard interactions remain computationally restricted to light or closed-shell nuclei ($A \lesssim 12$)~\cite{wiringa2014nucleon,piarulli2023densities,lonardoni2017variational}. While softened interactions enable converged computations for heavier nuclei, they mandate consistently evolved operators to preserve SRC physics. Consequently, most {\it ab initio} calculations of $\rho_2$ are restricted to light systems or use two-body operators~\cite{neff2015short,tropiano2024high,friman2021role}, even when the Hamiltonian includes vital three-body (3b) forces~\cite{Jurgenson2009SRGManyBodyForces,jurgenson2011evolving,roth2014evolved,hebeler2021three,miyagi2022converged}. Explicit treatment of induced 3b operator components has so far only been achieved for $^4$He~\cite{schuster2014operator}.

Recent experimental advances with inverse-kinematics reactions and radioactive beams now enable the probing of SRCs in mid-mass and open-shell nuclei, including unstable isotopes~\cite{patsyuk2021unperturbed,ye2024new}. This progress creates a pressing need for consistent theoretical predictions across the nuclear chart. In particular, a consistent treatment of 3b effects in both the Hamiltonian and operators is expected to grow in importance with increasing mass number, as the number of spectator nucleons rises. Yet the impact of consistently evolved 3b operators on SRC observables in medium-mass nuclei remains unexplored.

In this Letter, we present the first consistent study of 3b effects on SRCs using the in-medium similarity renormalization group (IMSRG)~\cite{tsukiyama2011medium,hergert2016medium,tsukiyama2012medium}. We evolve both the Hamiltonian and the density operator to low resolution via the free-space similarity renormalization group (SRG)~\cite{cruz2018short}, explicitly retaining all induced terms up to the 3b level. This consistent {\it ab initio} framework enables the calculation of $\rho_2$ and the derived nuclear contacts for a broad range of nuclei. We demonstrate that induced 3b operator contributions, while negligible in $^4\mathrm{He}$, become systematic and significant in heavier systems, revealing a previously missing piece in the microscopic description of SRCs across the nuclear chart.

\emph{Method.}---
Our approach adopts the IMSRG framework with consistently SRG-evolved input. The initial Hamiltonian is taken from chiral effective field theory (EFT):
\begin{equation}
  H = T_{\mathrm{int}} + V_{\rm NN} + V_{\rm 3N},
\end{equation}
where \(T_{\mathrm{int}}\) is the intrinsic kinetic energy, and \(V_{\rm NN}\) and \(V_{\rm 3N}\) are the bare two- and three-nucleon interactions. The target observable is the two-nucleon density, defined in coordinate $r$ and momentum $k$ space as
\begin{equation}
\begin{aligned}
\rho_{2}(r) &= \frac{1}{4 \pi r^2} \Big\langle \Psi \Big| \sum_{i<j}^A \delta\bigl(r-|\boldsymbol{r}_i-\boldsymbol{r}_j|\bigr) \Big| \Psi \Big\rangle, \\[4pt]
\rho_{2}(k) &= \frac{1}{4 \pi k^2} \Big\langle \Psi \Big| \sum_{i<j}^A \delta\bigl(k-|\boldsymbol{k}_i-\boldsymbol{k}_j|\bigr) \Big| \Psi \Big\rangle,
\end{aligned}
\end{equation}
normalized to \(A(A-1)/2\), where $A$ is mass number.

We first soften the interaction via free-space SRG evolution, governed by the flow equation \(dH_s/ds = [\eta_s, H_s]\) \cite{bogner2007similarity,bogner2010low,xu2024progress} with generator \(\eta^{\mathrm{SRG}}_s = [T_{\rm int}, H_s]\). 
The two-nucleon density operator is evolved consistently as $\rho_{2,s}=U_s \rho_2 U_s^{\dagger}$, which induces many-body terms 
\begin{equation}
\rho_{2,s}=\rho_{2}^{[2]}+\rho_{2}^{[3]}+\cdots.   
\end{equation}
To quantify truncation errors, we perform two sets of evolutions: 
\begin{itemize}
  \item \(H^{(2)}\) and \(\rho^{(2)}_2\): evolved in the two-nucleon (\(A=2\)) space using only the NN interaction; the evolved density operator is truncated at 2b level
  \begin{equation}
  \rho^{(2)}_2\equiv\rho_{2}^{[2]}.
  \end{equation}
  \item \(H^{(3)}\) and \(\rho^{(3)}_2\): evolved in the three-nucleon (\(A=3\)) space including both NN and 3N interactions; ; the evolved density operator is retained up to the three-body level,
  \begin{equation}
  \rho^{(3)}_2 \equiv \rho_{2}^{[2]}+\rho_{2}^{[3]}\,.
  \end{equation}
\end{itemize}
The evolution is halted at a finite resolution scale \(\lambda = s^{-1/4}\), which balances the decoupling of momentum modes against the growth of induced many-body forces. 

For the nuclear many-body calculation, we normal-order the SRG-evolved Hamiltonian with respect to a Hartree–Fock reference state \(|\Phi\rangle\), adopting the normal-ordered two-body (NO2B) approximation~\cite{roth2011similarity,roth2012medium}. This procedure absorbs dominant contributions from three-body forces into zero-, one-, and two-body terms. The subsequent IMSRG flow uses the White generator~\cite{white2002numerical} to decouple the reference state (for closed-shell nuclei) or a valence space (for open-shell nuclei). The resulting valence-space effective Hamiltonian \(H_{\mathrm{eff}}\) is diagonalized, and observables are computed using the consistently transformed operator \(O_{\mathrm{eff}}\).

The complete transformation for an expectation value is
\begin{equation}
\langle \Phi |\, \mathcal U\, U\, O\, U^\dagger \mathcal U^\dagger \,| \Phi \rangle,
\end{equation}
where \(U\) and \(\mathcal{U}\) are the free-space SRG and IMSRG unitaries, respectively, obtained by solving the flow equation. This expression is equivalent to \(\langle\Psi|O|\Psi\rangle\) with the correlated state \(|\Psi\rangle = U^\dagger\mathcal U^\dagger|\Phi\rangle\), thereby incorporating both short-range correlations (via SRG) and mid/long-range many-body correlations (via IMSRG) in a consistent framework.

\emph{Hamiltonian and parameters.}---
We present the two nucleon densities for $^4\mathrm{He}$ and oxygen isotopes calcluated with SRG-evolved operators. 
We use the chiral NN interaction at N$^4$LO of Ref.~\cite{entem2017high} together with the 3N force using local and nonlocal regulators, whose low-energy couplings $(c_D,c_E)$ are constrained to the triton half-life and binding energy~\cite{gysbers2019discrepancy}.
The bare Hamiltonians and operators are evolved in free space to several resolution scales $\lambda$ using \textsc{NuHamil}~\cite{miyagi2023nuhamil} in a relative-coordinate harmonic-oscillator (HO) basis.
Induced 3b terms are obtained via the subtraction method~\cite{roth2014evolved,schuster2014operator}, enabling a body-rank analysis.
The 3b evolution employed frequency conversion~\cite{roth2014evolved} from $\hbar\omega = 36$ MeV to 20 MeV in the ramp-C space~\cite{roth2014evolved}, which is sufficient achieve convergence for the oxygen isotopes.

The SRG-evolved Hamiltonians and two nucleon density operators are evolved in medium with \textsc{imsrg++}~\cite{StrobergIMSRG2021}.
The flow equations are solved in a HO basis with $\hbar\omega = 20$ MeV, truncating single-particle excitations at $e_{\mathrm{max}}=10$. 
And the 3b Hamiltonians and operators are truncated at $E_{3\mathrm{max}}=12$, where $E_{3\mathrm{max}} = \text{max}(e_1 +e_2 +e_3)$.
For closed-shell nuclei, the IMSRG(2) flow yields the ground-state energy and expectation values directly.
For open-shell nuclei, we construct the $H_{\mathrm{eff}}$ and diagonalize it with \textsc{KSHELL}~\cite{shimizu2019thick}.

\begin{figure}[t]
  \includegraphics[width=\columnwidth]{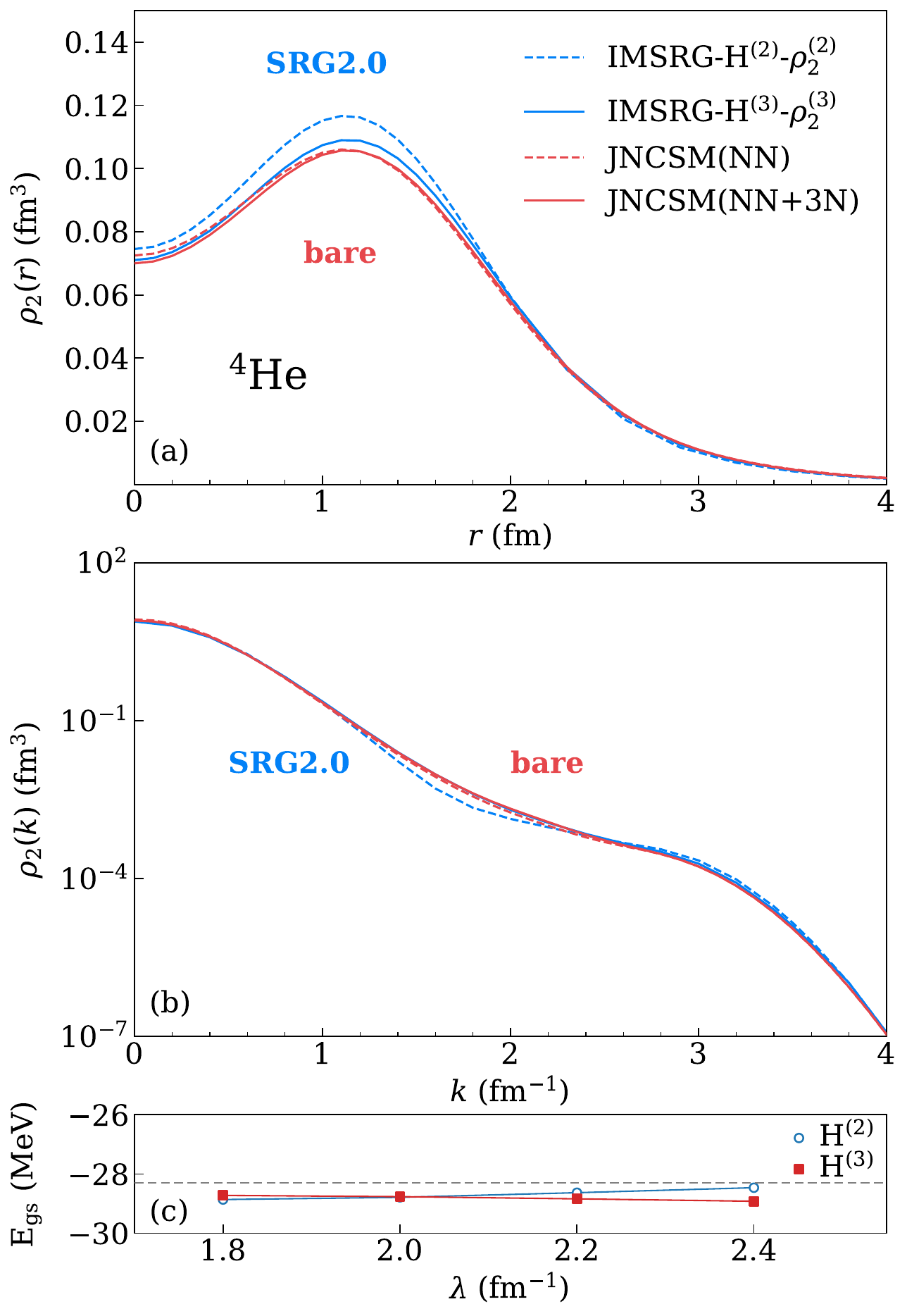}
  \caption{\label{fig:He4benchmark}
  (a) Coordinate-space two-nucleon density $\rho_2(r)$, (b) momentum-space distribution $\rho_2(k)$, and (c) ground-state energy of $^4\mathrm{He}$, calculated with the chiral NN-N$^4$LO+3$\mathrm{N}_{\mathrm{lnl}}$ interaction. 
  Blue curves are IMSRG results with consistently SRG-evolved Hamiltonians: dashed lines correspond to a two-body truncation ($H^{(2)},O^{(2)}$), and solid lines include full three-body terms ($H^{(3)},O^{(3)}$).. 
  Red curves are J-NCSM benchmark results obtained with the unevolved (bare) interaction: dashed for NN only, solid for NN+3N. 
  Panel (c) shows the ground-state energy as a function of the SRG resolution scale $\lambda$; the dashed horizontal line denotes the experimental value.
  }
\end{figure}

\emph{Results.}---
Figure~\ref{fig:He4benchmark}(a) and (b) show the coordinate- and momentum-space two-nucleon densities of $^4\mathrm{He}$, respectively. 
Due to its light mass, we can directly benchmark our operator-evolution results directly with 
exact diagonalizations of the four-body problem using no-core shell model in a Jacobi HO basis (J-NCSM)~\cite{barrett2013ab} with unevolved (bare) Hamiltonian and operators. 
These calculations use an oscillator frequency of $\hbar \omega =36$ MeV and a model space truncated at $N_{\mathrm{max}}=18$, which ensures convergence. 
For consistency, Figure~\ref{fig:He4benchmark} shows only rank-consistent cases where the Hamiltonian and operators are truncated at the same body rank. As demonstrated by the J-NCSM results, the NN-only and NN+3N density calculations are in close agreement, reflecting the minor role of three-nucleon forces in $^4\mathrm{He}$.


\begin{figure}[t]
  \includegraphics[width=\columnwidth]{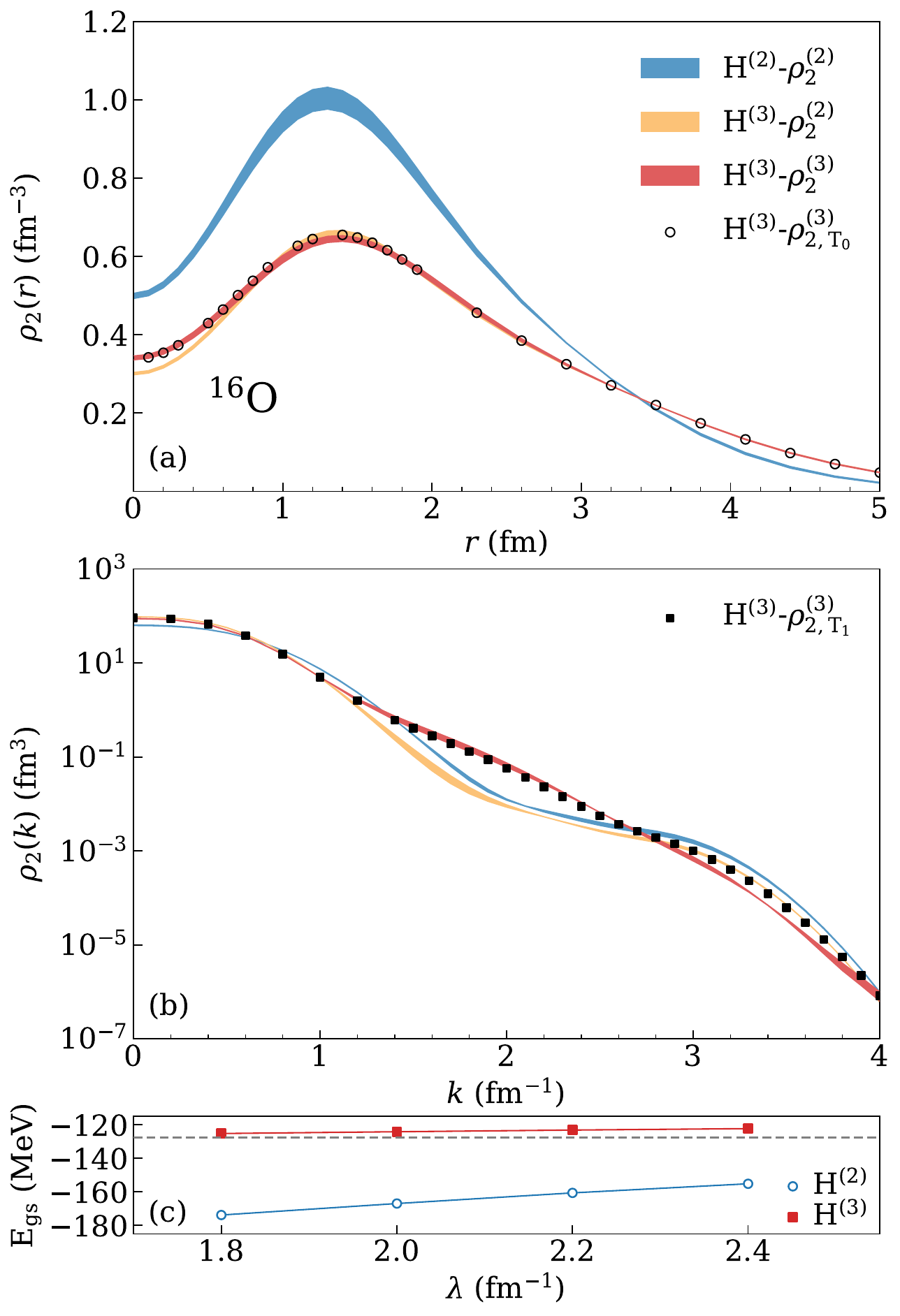}
  \caption{\label{fig:O16predict}
(a) Coordinate-space two-nucleon density $\rho_2(r)$, (b) momentum-space distribution $\rho_2(k)$, and (c) ground-state energy of $^{16}\mathrm{O}$, calculated with the chiral NN-N$^4$LO+3$\mathrm{N}_{\mathrm{lnl}}$ interaction.
All results are based on free-space SRG evolution, with bands indicating the variation over the resolution scale $\lambda=2.0$-2.4 fm$^{-1}$. 
Color denotes the truncation scheme: blue for the two-body truncation ($H^{(2)},O^{(2)}$); orange with three-body terms included only in the Hamiltonian ($H^{(3)},O^{(2)}$); and red with consistent three-body terms in both the Hamiltonian and the operators ($H^{(3)},O^{(3)}$). 
In coordinate space, the short-range part is dominated by $T=0$ (open circle), while in momentum space the intermediate- and high-momentum tail is dominated by $T=1$ (solid square).
Panel (c) shows $E_{\mathrm{gs}}$ versus $\lambda$ and the dashed line denotes the experimental value; including induced 3N terms reduces the $\lambda$ dependence and improves agreement with experiment. }
\end{figure}

Meanwhile, for the IMSRG calculations of $^4$He, the two-body truncation already reproduces the bulk features of $\rho_2$.  
Residual three-body effects are subdominant and adjust details: they suppress the short-distance part in (a) and enhance the medium/high-momentum tail in (b) 
around $k\!\sim\!2~\mathrm{fm}^{-1}$. 
At the 3b level, our results benchmark very well against the J-NCSM,  
indicating that SRG-evolved operators evaluated on IMSRG wave functions can reproduce density distributions obtained with bare 
interactions—supporting extensions to heavier and open-shell systems.

We also examine the dependence of the two-nucleon density $\rho_2$ and ground-state energy $E_{\rm gs}$ on the SRG resolution scale $\lambda$. For $^4$He, $\rho_2$ is virtually insensitive to $\lambda$ across the tested range of 1.8–2.4 fm$^{-1}$. As shown in Figure~\ref{fig:He4benchmark}(c), the ground-state energy exhibits a similarly weak dependence on $\lambda$ for both $H^{(2)}$ and $H^{(3)}$.

Moving to the heavier $^{16}\mathrm{O}$ system in Figure~\ref{fig:O16predict}, where exact diagonalization is intractable, we assess the three-body effects by evaluating SRG-evolved density operators on IMSRG(2) ground states.  
Panels (a) and (b) display the resulting coordinate- and momentum-space two-nucleon densities. 
As panel (c) indicates, the resolution-scale dependence remains significant when using only the two-body Hamiltonian $H^{(2)}$; we therefore represent these results as bands spanning $\lambda = 2.0$–$2.4~\mathrm{fm}^{-1}$.
Including the 3b Hamiltonian visibly narrows these bands, consistent with the reduction of $\lambda$ dependence in ground state energies. 
However, even with the full 3b treatment ($H^{(3)}, O^{(3)}$), a residual $\lambda$ dependence remains in $\rho_2$, reflecting induced higher-body terms beyond our truncations. 
Consequently, the width of these bands provides a practical proxy for the associated truncation uncertainty.

Compared with $^4\mathrm{He}$, the 3b corrections in $^{16}\mathrm{O}$ are significantly more pronounced. 
To isolate the individual contributions from the 3b Hamiltonian and the induced 3b operators, we compute three distinct truncation schemes: ($H^{(2)},O^{(2)}$), ($H^{(3)},O^{(2)}$) and ($H^{(3)},O^{(3)}$).
The Hamiltonian contribution is defined as $\Delta_H=(H^{(2)},O^{(2)})-(H^{(3)},O^{(2)})$, while the operator contribution is $\Delta_O=(H^{(3)},O^{(3)})-(H^{(3)},O^{(2)})$. 
Interestingly, we find that $\Delta_H$ mainly suppresses $\rho_2(r)$ at short distances and has only a small impact on the high-$k$ tail of $\rho_2(k)$, consistent with the repulsion feature in the 3b Hamiltonian at our resolution.  
In contrast, $\Delta_O$ primarily modifies the medium/high-momentum tail of $\rho_2(k)$ (around $k\approx2$ fm$^{-1}$) and has a minor effect on $\rho_2(r)$.

To further analyze this behavior, we separate the 3b operator correction into its isospin components. 
The open circles and solid squares denote results where only the $T=0$ or $T=1$ 3b correction is added to the two-body truncation, respectively (i.e., $\rho_2^{[2]} + \rho_{2,T}^{[3]}$ for $T = 0$ or $1$, where superscripts indicate the operator body rank). 
This analysis reveals a distinct isospin dependence: in momentum space, the $T=1$ ($pp/nn$) corrections are dominant, whereas in coordinate space, the short-range part is governed by the $T=0$ ($pn$) component.

This behavior can be understood through the mechanism of the SRG evolution. The generator $\eta^{\mathrm{SRG}}$ suppresses off-diagonal two-body couplings, the strongest of which involve tensor mixing. 
The strength removed from the two-body sector reappears in the form of  induced 3b operator; consequently, tensor interactions make a substantial contribution to these induced terms. 
In coordinate space, short-range region is dominated by $S$-wave ($L=0$), where the tensor terms mainly contributes in the $T=0$, $^3S_1$ channel. This explains why the induced $T=0$ 3b terms govern the short-range part of $\rho_2(r)$.
In momentum space, the medium-$k$ region (around $k \approx 2$ fm$^{-1}$ ) is dominated by $L>0$ partial waves. In $T=0$ channel, the two-body tensor force already generates  a significant $^3D_1$ component, leaving a more modest role for additional 3b corrections. 
By contrast, the dominant $T=1$ channel is the $^1S_0$ state at the two-body level. Here, the induced 3b operator provides the leading $L>0$ components, which in turn dominate the medium-$k$ behavior of $\rho_2(k)$~\cite{neff2015short,feldmeier2011universality}.

 
\begin{figure}[t]
  \includegraphics[width=\columnwidth]{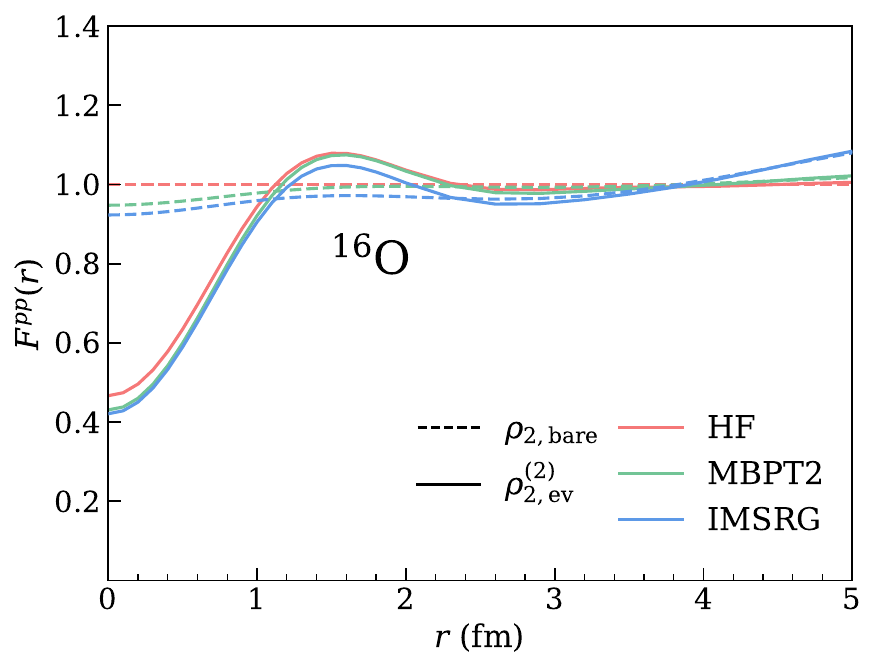}
  \caption{\label{fig:O16corr_pp}
  Correlation function $F^{pp}(r)$ of $^{16}\mathrm{O}$ defined in Eq.~\eqref{eq:corfun}. 
The numerator $\rho_2$ is calculated using either the bare (dashed lines) or SRG-evolved (solid lines) two-body operator, applied to states obtained from HF (red), second-order many-body perturbation theory (MBPT2, orange), and the IMSRG (blue). 
The denominator is the HF two nucleon density evaluated with the bare operator. 
  }
\end{figure}

The combined SRG-IMSRG evolution of operators corresponds to generating correlations starting from the Hartree-Fock (HF) state, expressed as $|\Psi_{s=0}\rangle=U^{\dagger}\mathcal{U}^{\dagger}|\Phi\rangle$. To quantify these correlations, we compute the $pp$ correlation function defined as~\cite{cruz2018short}
\begin{equation}
  F^{pp}(r) \equiv \frac{\rho^{pp}(r)}{\rho^{pp}_{\text {uncorr}}(r)} , \label{eq:corfun}
\end{equation}
where the denominator $\rho^{pp}_{\text{uncorr}}(r)$ is the uncorrelated two-nucleon density computed from the HF state of the SRG-evolved Hamiltonian using the bare operator.
For the numerator, we first evaluate $\rho^{pp}(r)$ with the same bare operator on many-body states of increasing sophistication—HF, second-order many-body perturbation theory (MBPT2), and IMSRG—all constructed from the softened Hamiltonian. Conesequently, for all the approaches, the deviation of $F^{pp}(r)$ from unity remains small, as shown by the dashed lines in Fig.~\ref{fig:O16corr_pp}. 
We then replace the operators by thier SRG-evolved form (here truncated at two-body order in the $T=1$ channel, where 3b corrections are small) while keeping the many-body states fixed. 
This evolution leaves $F^{pp}(r)$ essentially unchanged at large distances ($r\gtrsim 2$-$3$ fm), but induces a pronounced enhancement at short-to-intermediate distances ($r\lesssim 2$ fm). This result demonstrates that the IMSRG captures most of the long‑range (large‑$r$) correlations, while the free‑space SRG evolution of the operator supplies the missing short‑range strength.

\begin{figure}[t]
  \includegraphics[width=\columnwidth]{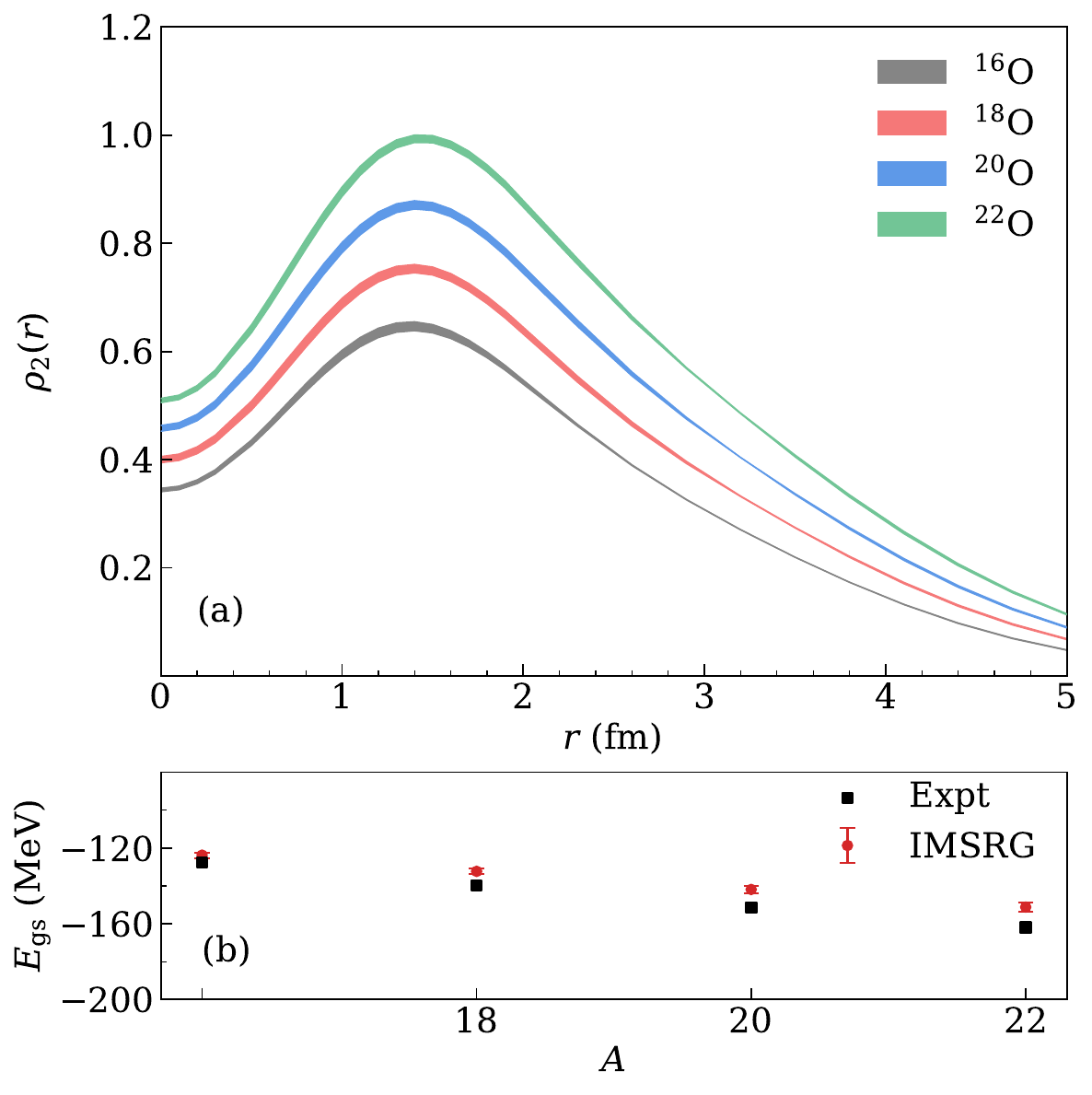}
  \caption{\label{fig:Oiso}
  (a) Cordinate-space two-nucleon densitises and (b) ground state energies of oxygen isotopes. 
  Calculations use constantly evolved Hamiltonians and operators truncated to 3b.
  The bands in (a) and error bars in (b) indicate the variation over the resolution scale $\lambda=2.0$-2.4 fm$^{-1}$. 
  }
\end{figure}

Furthermore, we extend our studies to oxygen isotopes,  encompassing both closed- and open-shell nuclei. Figure~\ref{fig:Oiso} presents the resulting two-nucleon densities and ground-state energies. Panel (a) displays the coordinate-space $\rho_2(r)$ evaluated with SRG-evolved operators truncated at 3b level on IMSRG (closed shell) and VS-IMSRG (open shell) states. 
For $r\!\lesssim\!1.5$ fm the curves exhibit a common trend across isotopes, reflecting the universal short-range repulsion~\cite{cruz2021many}. 
Panel (b) shows the corresponding ground-state energies, whichare in good agreement with experimental data. 
Together, these panels provide a consistent \textit{ab initio} description of low-energy structure and SRCs  across medium-mass nuclei.

\emph{Summary.}---
In this Letter we present \textit{ab initio} calculations of two-nucleon densities to study SRCs in medium-mass nuclei. Our approach combines free-space SRG-evolved density operators with IMSRG many-body states. 
Benchmark results in $^4\mathrm{He}$ validate the framework. 
The key advance is the consistent inclusion of SRG-induced 3b operator components, which together with the 3b Hamiltonian provides a complete treatment of 3b effects.
These effects are pronuouced in $^{16}\mathrm{O}$: 
the 3b Hamiltonian mainly suppresses the short-range part of $\rho_2(r)$, whereas the 3b operator corrects the high-momentum tail of $\rho_2(k)$.
These 3b contributions may be linked to recently extracted three-nucleon SRCs in Ref.~\cite{li2025inclusive}.

For the oxygen isotopes, our framework provides a consistent description of both low-energy structure and SRCs in medium-mass nuclei. 
In nuclei without direct SRC measurements, we can validate the calculations by reproducing available low-energy observables and subsequently predict the SRC-sensitive two-nucleon densities. This predictive capability is valuable for clarifying the reported suppression of SRC strength in neutron-rich systems~\cite{hen2014momentum,nguyen2020novel} and for guiding future inverse-kinematics measurements of SRCs in unstable nuclei~\cite{patsyuk2021unperturbed,ye2024new}.

\emph{Acknowledgements.}--- 
This material is based upon work supported by the National Natural Science Foundation of China under Contract No.\,12347106 and No.\,12547102; the National Key Research and Development Program (MOST 2023YFA1606404 and MOST 2022YFA1602303). 

\emph{Data availability.}---
The data are not publicly available.
The data are available from the authors upon reasonable
request.

\bibliography{ref} 

\begin{thebibliography}{48}%
\makeatletter
\providecommand \@ifxundefined [1]{%
 \@ifx{#1\undefined}
}%
\providecommand \@ifnum [1]{%
 \ifnum #1\expandafter \@firstoftwo
 \else \expandafter \@secondoftwo
 \fi
}%
\providecommand \@ifx [1]{%
 \ifx #1\expandafter \@firstoftwo
 \else \expandafter \@secondoftwo
 \fi
}%
\providecommand \natexlab [1]{#1}%
\providecommand \enquote  [1]{``#1''}%
\providecommand \bibnamefont  [1]{#1}%
\providecommand \bibfnamefont [1]{#1}%
\providecommand \citenamefont [1]{#1}%
\providecommand \href@noop [0]{\@secondoftwo}%
\providecommand \href [0]{\begingroup \@sanitize@url \@href}%
\providecommand \@href[1]{\@@startlink{#1}\@@href}%
\providecommand \@@href[1]{\endgroup#1\@@endlink}%
\providecommand \@sanitize@url [0]{\catcode `\\12\catcode `\$12\catcode
  `\&12\catcode `\#12\catcode `\^12\catcode `\_12\catcode `\%12\relax}%
\providecommand \@@startlink[1]{}%
\providecommand \@@endlink[0]{}%
\providecommand \url  [0]{\begingroup\@sanitize@url \@url }%
\providecommand \@url [1]{\endgroup\@href {#1}{\urlprefix }}%
\providecommand \urlprefix  [0]{URL }%
\providecommand \Eprint [0]{\href }%
\providecommand \doibase [0]{https://doi.org/}%
\providecommand \selectlanguage [0]{\@gobble}%
\providecommand \bibinfo  [0]{\@secondoftwo}%
\providecommand \bibfield  [0]{\@secondoftwo}%
\providecommand \translation [1]{[#1]}%
\providecommand \BibitemOpen [0]{}%
\providecommand \bibitemStop [0]{}%
\providecommand \bibitemNoStop [0]{.\EOS\space}%
\providecommand \EOS [0]{\spacefactor3000\relax}%
\providecommand \BibitemShut  [1]{\csname bibitem#1\endcsname}%
\let\auto@bib@innerbib\@empty
\bibitem [{\citenamefont {Hen}\ \emph {et~al.}(2017)\citenamefont {Hen},
  \citenamefont {Miller}, \citenamefont {Piasetzky},\ and\ \citenamefont
  {Weinstein}}]{hen2017nucleon}%
  \BibitemOpen
  \bibfield  {author} {\bibinfo {author} {\bibfnamefont {O.}~\bibnamefont
  {Hen}}, \bibinfo {author} {\bibfnamefont {G.~A.}\ \bibnamefont {Miller}},
  \bibinfo {author} {\bibfnamefont {E.}~\bibnamefont {Piasetzky}},\ and\
  \bibinfo {author} {\bibfnamefont {L.~B.}\ \bibnamefont {Weinstein}},\
  }\bibfield  {title} {\bibinfo {title} {{Nucleon-nucleon correlations,
  short-lived excitations, and the quarks within}},\ }\href
  {https://doi.org/10.1103/RevModPhys.89.045002} {\bibfield  {journal}
  {\bibinfo  {journal} {Rev. Mod. Phys.}\ }\textbf {\bibinfo {volume} {89}},\
  \bibinfo {pages} {045002} (\bibinfo {year} {2017})}\BibitemShut {NoStop}%
\bibitem [{\citenamefont {Schmidt}\ \emph {et~al.}(2020)\citenamefont
  {Schmidt}, \citenamefont {Pybus}, \citenamefont {Weiss}, \citenamefont
  {Segarra}, \citenamefont {Hrnjic}, \citenamefont {Denniston}, \citenamefont
  {Hen}, \citenamefont {Piasetzky}, \citenamefont {Weinstein}, \citenamefont
  {Barnea} \emph {et~al.}}]{schmidt2020probing}%
  \BibitemOpen
  \bibfield  {author} {\bibinfo {author} {\bibfnamefont {A.}~\bibnamefont
  {Schmidt}}, \bibinfo {author} {\bibfnamefont {J.~R.}\ \bibnamefont {Pybus}},
  \bibinfo {author} {\bibfnamefont {R.}~\bibnamefont {Weiss}}, \bibinfo
  {author} {\bibfnamefont {E.~P.}\ \bibnamefont {Segarra}}, \bibinfo {author}
  {\bibfnamefont {A.}~\bibnamefont {Hrnjic}}, \bibinfo {author} {\bibfnamefont
  {A.}~\bibnamefont {Denniston}}, \bibinfo {author} {\bibfnamefont
  {O.}~\bibnamefont {Hen}}, \bibinfo {author} {\bibfnamefont {E.}~\bibnamefont
  {Piasetzky}}, \bibinfo {author} {\bibfnamefont {L.~B.}\ \bibnamefont
  {Weinstein}}, \bibinfo {author} {\bibfnamefont {N.}~\bibnamefont {Barnea}},
  \emph {et~al.},\ }\bibfield  {title} {\bibinfo {title} {{Probing the core of
  the strong nuclear interaction}},\ }\href
  {https://doi.org/10.1038/s41586-020-2021-6} {\bibfield  {journal} {\bibinfo
  {journal} {Nature}\ }\textbf {\bibinfo {volume} {578}},\ \bibinfo {pages}
  {540} (\bibinfo {year} {2020})}\BibitemShut {NoStop}%
\bibitem [{\citenamefont {Korover}\ \emph {et~al.}(2014)\citenamefont
  {Korover}, \citenamefont {Muangma}, \citenamefont {Hen}, \citenamefont
  {Shneor}, \citenamefont {Sulkosky}, \citenamefont {Kelleher}, \citenamefont
  {Gilad}, \citenamefont {Higinbotham}, \citenamefont {Piasetzky},
  \citenamefont {Watson} \emph {et~al.}}]{korover2014probing}%
  \BibitemOpen
  \bibfield  {author} {\bibinfo {author} {\bibfnamefont {I.}~\bibnamefont
  {Korover}}, \bibinfo {author} {\bibfnamefont {N.}~\bibnamefont {Muangma}},
  \bibinfo {author} {\bibfnamefont {O.}~\bibnamefont {Hen}}, \bibinfo {author}
  {\bibfnamefont {R.}~\bibnamefont {Shneor}}, \bibinfo {author} {\bibfnamefont
  {V.}~\bibnamefont {Sulkosky}}, \bibinfo {author} {\bibfnamefont
  {A.}~\bibnamefont {Kelleher}}, \bibinfo {author} {\bibfnamefont
  {S.}~\bibnamefont {Gilad}}, \bibinfo {author} {\bibfnamefont {D.~W.}\
  \bibnamefont {Higinbotham}}, \bibinfo {author} {\bibfnamefont
  {E.}~\bibnamefont {Piasetzky}}, \bibinfo {author} {\bibfnamefont {J.~W.}\
  \bibnamefont {Watson}}, \emph {et~al.},\ }\bibfield  {title} {\bibinfo
  {title} {{Probing the repulsive core of the nucleon-nucleon interaction via
  the $^{4}\mathrm{He} (e, e^{\prime}p N)$ triple-coincidence reaction}},\
  }\href {https://doi.org/10.1103/PhysRevLett.113.022501} {\bibfield  {journal}
  {\bibinfo  {journal} {Phys. Rev. Lett.}\ }\textbf {\bibinfo {volume} {113}},\
  \bibinfo {pages} {022501} (\bibinfo {year} {2014})}\BibitemShut {NoStop}%
\bibitem [{\citenamefont {Cai}\ \emph {et~al.}(2025)\citenamefont {Cai},
  \citenamefont {Li},\ and\ \citenamefont {Ma}}]{cai2025srceosdensematter}%
  \BibitemOpen
  \bibfield  {author} {\bibinfo {author} {\bibfnamefont {B.}~\bibnamefont
  {Cai}}, \bibinfo {author} {\bibfnamefont {B.}~\bibnamefont {Li}},\ and\
  \bibinfo {author} {\bibfnamefont {Y.}~\bibnamefont {Ma}},\ }\bibfield
  {title} {\bibinfo {title} {{Nucleon Short-Range Correlations and
  High-Momentum Dynamics: Implications on the Equation of State of Dense
  Matter}},\ }\bibfield  {journal} {\bibinfo  {journal} {arXiv}\ }\href
  {https://doi.org/10.48550/arXiv.2512.04206} {10.48550/arXiv.2512.04206}
  (\bibinfo {year} {2025})\BibitemShut {NoStop}%
\bibitem [{\citenamefont {Denniston}\ \emph {et~al.}(2024)\citenamefont
  {Denniston}, \citenamefont {Je{\v{z}}o}, \citenamefont {Kusina},
  \citenamefont {Derakhshanian}, \citenamefont {Duwent{\"a}ster}, \citenamefont
  {Hen}, \citenamefont {Keppel}, \citenamefont {Klasen}, \citenamefont
  {Kova{\v{r}}{\'\i}k}, \citenamefont {Morf{\'\i}n} \emph
  {et~al.}}]{denniston2024modification}%
  \BibitemOpen
  \bibfield  {author} {\bibinfo {author} {\bibfnamefont {A.~W.}\ \bibnamefont
  {Denniston}}, \bibinfo {author} {\bibfnamefont {T.}~\bibnamefont
  {Je{\v{z}}o}}, \bibinfo {author} {\bibfnamefont {A.}~\bibnamefont {Kusina}},
  \bibinfo {author} {\bibfnamefont {N.}~\bibnamefont {Derakhshanian}}, \bibinfo
  {author} {\bibfnamefont {P.}~\bibnamefont {Duwent{\"a}ster}}, \bibinfo
  {author} {\bibfnamefont {O.}~\bibnamefont {Hen}}, \bibinfo {author}
  {\bibfnamefont {C.}~\bibnamefont {Keppel}}, \bibinfo {author} {\bibfnamefont
  {M.}~\bibnamefont {Klasen}}, \bibinfo {author} {\bibfnamefont
  {K.}~\bibnamefont {Kova{\v{r}}{\'\i}k}}, \bibinfo {author} {\bibfnamefont
  {J.~G.}\ \bibnamefont {Morf{\'\i}n}}, \emph {et~al.},\ }\bibfield  {title}
  {\bibinfo {title} {{Modification of quark-gluon distributions in nuclei by
  correlated nucleon pairs}},\ }\href
  {https://doi.org/10.1103/PhysRevLett.133.152502} {\bibfield  {journal}
  {\bibinfo  {journal} {Phys. Rev. Lett.}\ }\textbf {\bibinfo {volume} {133}},\
  \bibinfo {pages} {152502} (\bibinfo {year} {2024})}\BibitemShut {NoStop}%
\bibitem [{\citenamefont {Weinstein}\ \emph {et~al.}(2011)\citenamefont
  {Weinstein}, \citenamefont {Piasetzky}, \citenamefont {Higinbotham},
  \citenamefont {Gomez}, \citenamefont {Hen},\ and\ \citenamefont
  {Shneor}}]{weinstein2011short}%
  \BibitemOpen
  \bibfield  {author} {\bibinfo {author} {\bibfnamefont {L.~B.}\ \bibnamefont
  {Weinstein}}, \bibinfo {author} {\bibfnamefont {E.}~\bibnamefont
  {Piasetzky}}, \bibinfo {author} {\bibfnamefont {D.~W.}\ \bibnamefont
  {Higinbotham}}, \bibinfo {author} {\bibfnamefont {J.}~\bibnamefont {Gomez}},
  \bibinfo {author} {\bibfnamefont {O.}~\bibnamefont {Hen}},\ and\ \bibinfo
  {author} {\bibfnamefont {R.}~\bibnamefont {Shneor}},\ }\bibfield  {title}
  {\bibinfo {title} {{Short range correlations and the EMC effect}},\ }\href
  {https://doi.org/10.1103/PhysRevLett.106.052301} {\bibfield  {journal}
  {\bibinfo  {journal} {Phys. Rev. Lett.}\ }\textbf {\bibinfo {volume} {106}},\
  \bibinfo {pages} {052301} (\bibinfo {year} {2011})}\BibitemShut {NoStop}%
\bibitem [{\citenamefont {degli Atti}(2015)}]{degli2015medium}%
  \BibitemOpen
  \bibfield  {author} {\bibinfo {author} {\bibfnamefont {C.~C.}\ \bibnamefont
  {degli Atti}},\ }\bibfield  {title} {\bibinfo {title} {{In-medium short-range
  dynamics of nucleons: Recent theoretical and experimental advances}},\ }\href
  {https://doi.org/10.1016/j.physrep.2015.06.002} {\bibfield  {journal}
  {\bibinfo  {journal} {Phys. Rep.}\ }\textbf {\bibinfo {volume} {590}},\
  \bibinfo {pages} {1} (\bibinfo {year} {2015})}\BibitemShut {NoStop}%
\bibitem [{\citenamefont {Chen}\ \emph {et~al.}(2017)\citenamefont {Chen},
  \citenamefont {Detmold}, \citenamefont {Lynn},\ and\ \citenamefont
  {Schwenk}}]{chen2017short}%
  \BibitemOpen
  \bibfield  {author} {\bibinfo {author} {\bibfnamefont {J.-W.}\ \bibnamefont
  {Chen}}, \bibinfo {author} {\bibfnamefont {W.}~\bibnamefont {Detmold}},
  \bibinfo {author} {\bibfnamefont {J.~E.}\ \bibnamefont {Lynn}},\ and\
  \bibinfo {author} {\bibfnamefont {A.}~\bibnamefont {Schwenk}},\ }\bibfield
  {title} {\bibinfo {title} {{Short-range correlations and the EMC effect in
  effective field theory}},\ }\href
  {https://doi.org/10.1103/PhysRevLett.119.262502} {\bibfield  {journal}
  {\bibinfo  {journal} {Phys. Rev. Lett.}\ }\textbf {\bibinfo {volume} {119}},\
  \bibinfo {pages} {262502} (\bibinfo {year} {2017})}\BibitemShut {NoStop}%
\bibitem [{\citenamefont {Weiss}\ \emph {et~al.}(2015)\citenamefont {Weiss},
  \citenamefont {Bazak},\ and\ \citenamefont {Barnea}}]{weiss2015generalized}%
  \BibitemOpen
  \bibfield  {author} {\bibinfo {author} {\bibfnamefont {R.}~\bibnamefont
  {Weiss}}, \bibinfo {author} {\bibfnamefont {B.}~\bibnamefont {Bazak}},\ and\
  \bibinfo {author} {\bibfnamefont {N.}~\bibnamefont {Barnea}},\ }\bibfield
  {title} {\bibinfo {title} {{Generalized nuclear contacts and momentum
  distributions}},\ }\href {https://doi.org/10.1103/PhysRevC.92.054311}
  {\bibfield  {journal} {\bibinfo  {journal} {Phys. Rev. C}\ }\textbf {\bibinfo
  {volume} {92}},\ \bibinfo {pages} {054311} (\bibinfo {year}
  {2015})}\BibitemShut {NoStop}%
\bibitem [{\citenamefont {Cruz-Torres}\ \emph {et~al.}(2021)\citenamefont
  {Cruz-Torres}, \citenamefont {Lonardoni}, \citenamefont {Weiss},
  \citenamefont {Piarulli}, \citenamefont {Barnea}, \citenamefont
  {Higinbotham}, \citenamefont {Piasetzky}, \citenamefont {Schmidt},
  \citenamefont {Weinstein}, \citenamefont {Wiringa} \emph
  {et~al.}}]{cruz2021many}%
  \BibitemOpen
  \bibfield  {author} {\bibinfo {author} {\bibfnamefont {R.}~\bibnamefont
  {Cruz-Torres}}, \bibinfo {author} {\bibfnamefont {D.}~\bibnamefont
  {Lonardoni}}, \bibinfo {author} {\bibfnamefont {R.}~\bibnamefont {Weiss}},
  \bibinfo {author} {\bibfnamefont {M.}~\bibnamefont {Piarulli}}, \bibinfo
  {author} {\bibfnamefont {N.}~\bibnamefont {Barnea}}, \bibinfo {author}
  {\bibfnamefont {D.}~\bibnamefont {Higinbotham}}, \bibinfo {author}
  {\bibfnamefont {E.}~\bibnamefont {Piasetzky}}, \bibinfo {author}
  {\bibfnamefont {A.}~\bibnamefont {Schmidt}}, \bibinfo {author} {\bibfnamefont
  {L.}~\bibnamefont {Weinstein}}, \bibinfo {author} {\bibfnamefont
  {R.}~\bibnamefont {Wiringa}}, \emph {et~al.},\ }\bibfield  {title} {\bibinfo
  {title} {{Many-body factorization and position--momentum equivalence of
  nuclear short-range correlations}},\ }\href
  {https://doi.org/10.1038/s41567-020-01053-7} {\bibfield  {journal} {\bibinfo
  {journal} {Nat. Phys.}\ }\textbf {\bibinfo {volume} {17}},\ \bibinfo {pages}
  {306} (\bibinfo {year} {2021})}\BibitemShut {NoStop}%
\bibitem [{\citenamefont {Tropiano}\ \emph {et~al.}(2021)\citenamefont
  {Tropiano}, \citenamefont {Bogner},\ and\ \citenamefont
  {Furnstahl}}]{tropiano2021short}%
  \BibitemOpen
  \bibfield  {author} {\bibinfo {author} {\bibfnamefont {A.~J.}\ \bibnamefont
  {Tropiano}}, \bibinfo {author} {\bibfnamefont {S.~K.}\ \bibnamefont
  {Bogner}},\ and\ \bibinfo {author} {\bibfnamefont {R.~J.}\ \bibnamefont
  {Furnstahl}},\ }\bibfield  {title} {\bibinfo {title} {{Short-range
  correlation physics at low renormalization group resolution}},\ }\href
  {https://doi.org/10.1103/PhysRevC.104.034311} {\bibfield  {journal} {\bibinfo
   {journal} {Phys. Rev. C}\ }\textbf {\bibinfo {volume} {104}},\ \bibinfo
  {pages} {034311} (\bibinfo {year} {2021})}\BibitemShut {NoStop}%
\bibitem [{\citenamefont {Tropiano}\ \emph {et~al.}(2024)\citenamefont
  {Tropiano}, \citenamefont {Bogner}, \citenamefont {Furnstahl}, \citenamefont
  {Hisham}, \citenamefont {Lovato},\ and\ \citenamefont
  {Wiringa}}]{tropiano2024high}%
  \BibitemOpen
  \bibfield  {author} {\bibinfo {author} {\bibfnamefont {A.~J.}\ \bibnamefont
  {Tropiano}}, \bibinfo {author} {\bibfnamefont {S.~K.}\ \bibnamefont
  {Bogner}}, \bibinfo {author} {\bibfnamefont {R.~J.}\ \bibnamefont
  {Furnstahl}}, \bibinfo {author} {\bibfnamefont {M.~A.}\ \bibnamefont
  {Hisham}}, \bibinfo {author} {\bibfnamefont {A.}~\bibnamefont {Lovato}},\
  and\ \bibinfo {author} {\bibfnamefont {R.~B.}\ \bibnamefont {Wiringa}},\
  }\bibfield  {title} {\bibinfo {title} {{High-resolution momentum
  distributions from low-resolution wave functions}},\ }\href
  {https://doi.org/10.1016/j.physletb.2024.138591} {\bibfield  {journal}
  {\bibinfo  {journal} {Phys. Lett. B}\ }\textbf {\bibinfo {volume} {852}},\
  \bibinfo {pages} {138591} (\bibinfo {year} {2024})}\BibitemShut {NoStop}%
\bibitem [{\citenamefont {Shang}\ \emph {et~al.}(2025)\citenamefont {Shang},
  \citenamefont {Chen}, \citenamefont {Hu}, \citenamefont {Zhen}, \citenamefont
  {Jiang},\ and\ \citenamefont {Pei}}]{shang2025many}%
  \BibitemOpen
  \bibfield  {author} {\bibinfo {author} {\bibfnamefont {H.}~\bibnamefont
  {Shang}}, \bibinfo {author} {\bibfnamefont {J.}~\bibnamefont {Chen}},
  \bibinfo {author} {\bibfnamefont {R.}~\bibnamefont {Hu}}, \bibinfo {author}
  {\bibfnamefont {X.}~\bibnamefont {Zhen}}, \bibinfo {author} {\bibfnamefont
  {C.}~\bibnamefont {Jiang}},\ and\ \bibinfo {author} {\bibfnamefont
  {J.}~\bibnamefont {Pei}},\ }\bibfield  {title} {\bibinfo {title} {Many-body
  effects on nuclear short range correlations},\ }\href
  {https://doi.org/10.1016/j.physletb.2025.139976} {\bibfield  {journal}
  {\bibinfo  {journal} {Phys. Lett. B}\ }\textbf {\bibinfo {volume} {871}},\
  \bibinfo {pages} {139976} (\bibinfo {year} {2025})}\BibitemShut {NoStop}%
\bibitem [{\citenamefont {Bogner}\ \emph {et~al.}(2007)\citenamefont {Bogner},
  \citenamefont {Furnstahl},\ and\ \citenamefont
  {Perry}}]{bogner2007similarity}%
  \BibitemOpen
  \bibfield  {author} {\bibinfo {author} {\bibfnamefont {S.~K.}\ \bibnamefont
  {Bogner}}, \bibinfo {author} {\bibfnamefont {R.~J.}\ \bibnamefont
  {Furnstahl}},\ and\ \bibinfo {author} {\bibfnamefont {R.~J.}\ \bibnamefont
  {Perry}},\ }\bibfield  {title} {\bibinfo {title} {{Similarity renormalization
  group for nucleon-nucleon interactions}},\ }\href
  {https://doi.org/10.1103/PhysRevC.75.061001} {\bibfield  {journal} {\bibinfo
  {journal} {Phys. Rev. C}\ }\textbf {\bibinfo {volume} {75}},\ \bibinfo
  {pages} {061001} (\bibinfo {year} {2007})}\BibitemShut {NoStop}%
\bibitem [{\citenamefont {Bogner}\ \emph {et~al.}(2010)\citenamefont {Bogner},
  \citenamefont {Furnstahl},\ and\ \citenamefont {Schwenk}}]{bogner2010low}%
  \BibitemOpen
  \bibfield  {author} {\bibinfo {author} {\bibfnamefont {S.~K.}\ \bibnamefont
  {Bogner}}, \bibinfo {author} {\bibfnamefont {R.~J.}\ \bibnamefont
  {Furnstahl}},\ and\ \bibinfo {author} {\bibfnamefont {A.}~\bibnamefont
  {Schwenk}},\ }\bibfield  {title} {\bibinfo {title} {{From low-momentum
  interactions to nuclear structure}},\ }\href
  {https://doi.org/10.1016/j.ppnp.2010.03.001} {\bibfield  {journal} {\bibinfo
  {journal} {Prog. Part. Nucl. Phys.}\ }\textbf {\bibinfo {volume} {65}},\
  \bibinfo {pages} {94} (\bibinfo {year} {2010})}\BibitemShut {NoStop}%
\bibitem [{\citenamefont {Stetcu}\ \emph {et~al.}(2006)\citenamefont {Stetcu},
  \citenamefont {Barrett}, \citenamefont {Navr{\'a}til},\ and\ \citenamefont
  {Vary}}]{stetcu2006long}%
  \BibitemOpen
  \bibfield  {author} {\bibinfo {author} {\bibfnamefont {I.}~\bibnamefont
  {Stetcu}}, \bibinfo {author} {\bibfnamefont {B.~R.}\ \bibnamefont {Barrett}},
  \bibinfo {author} {\bibfnamefont {P.}~\bibnamefont {Navr{\'a}til}},\ and\
  \bibinfo {author} {\bibfnamefont {J.~P.}\ \bibnamefont {Vary}},\ }\bibfield
  {title} {\bibinfo {title} {{Long-and short-range correlations in the
  ab-initio no-core shell model}},\ }\href
  {https://doi.org/10.1103/PhysRevC.73.037307} {\bibfield  {journal} {\bibinfo
  {journal} {Phys. Rev. C}\ }\textbf {\bibinfo {volume} {73}},\ \bibinfo
  {pages} {037307} (\bibinfo {year} {2006})}\BibitemShut {NoStop}%
\bibitem [{\citenamefont {Anderson}\ \emph {et~al.}(2010)\citenamefont
  {Anderson}, \citenamefont {Bogner}, \citenamefont {Furnstahl},\ and\
  \citenamefont {Perry}}]{anderson2010operator}%
  \BibitemOpen
  \bibfield  {author} {\bibinfo {author} {\bibfnamefont {E.~R.}\ \bibnamefont
  {Anderson}}, \bibinfo {author} {\bibfnamefont {S.~K.}\ \bibnamefont
  {Bogner}}, \bibinfo {author} {\bibfnamefont {R.~J.}\ \bibnamefont
  {Furnstahl}},\ and\ \bibinfo {author} {\bibfnamefont {R.}~\bibnamefont
  {Perry}},\ }\bibfield  {title} {\bibinfo {title} {{Operator evolution via the
  similarity renormalization group: The deuteron}},\ }\href
  {https://doi.org/10.1103/PhysRevC.82.054001} {\bibfield  {journal} {\bibinfo
  {journal} {Phys. Rev. C}\ }\textbf {\bibinfo {volume} {82}},\ \bibinfo
  {pages} {054001} (\bibinfo {year} {2010})}\BibitemShut {NoStop}%
\bibitem [{\citenamefont {Neff}\ \emph {et~al.}(2015)\citenamefont {Neff},
  \citenamefont {Feldmeier},\ and\ \citenamefont {Horiuchi}}]{neff2015short}%
  \BibitemOpen
  \bibfield  {author} {\bibinfo {author} {\bibfnamefont {T.}~\bibnamefont
  {Neff}}, \bibinfo {author} {\bibfnamefont {H.}~\bibnamefont {Feldmeier}},\
  and\ \bibinfo {author} {\bibfnamefont {W.}~\bibnamefont {Horiuchi}},\
  }\bibfield  {title} {\bibinfo {title} {{Short-range correlations in nuclei
  with similarity renormalization group transformations}},\ }\href
  {https://doi.org/10.1103/PhysRevC.92.024003} {\bibfield  {journal} {\bibinfo
  {journal} {Phys. Rev. C}\ }\textbf {\bibinfo {volume} {92}},\ \bibinfo
  {pages} {024003} (\bibinfo {year} {2015})}\BibitemShut {NoStop}%
\bibitem [{\citenamefont {Wiringa}\ \emph {et~al.}(2014)\citenamefont
  {Wiringa}, \citenamefont {Schiavilla}, \citenamefont {Pieper},\ and\
  \citenamefont {Carlson}}]{wiringa2014nucleon}%
  \BibitemOpen
  \bibfield  {author} {\bibinfo {author} {\bibfnamefont {R.~B.}\ \bibnamefont
  {Wiringa}}, \bibinfo {author} {\bibfnamefont {R.}~\bibnamefont {Schiavilla}},
  \bibinfo {author} {\bibfnamefont {S.~C.}\ \bibnamefont {Pieper}},\ and\
  \bibinfo {author} {\bibfnamefont {J.}~\bibnamefont {Carlson}},\ }\bibfield
  {title} {\bibinfo {title} {{Nucleon and nucleon-pair momentum distributions
  in $A \le 12$ nuclei}},\ }\href {https://doi.org/10.1103/PhysRevC.89.024305}
  {\bibfield  {journal} {\bibinfo  {journal} {Phys. Rev. C}\ }\textbf {\bibinfo
  {volume} {89}},\ \bibinfo {pages} {024305} (\bibinfo {year}
  {2014})}\BibitemShut {NoStop}%
\bibitem [{\citenamefont {Piarulli}\ \emph {et~al.}(2023)\citenamefont
  {Piarulli}, \citenamefont {Pastore}, \citenamefont {Wiringa}, \citenamefont
  {Brusilow},\ and\ \citenamefont {Lim}}]{piarulli2023densities}%
  \BibitemOpen
  \bibfield  {author} {\bibinfo {author} {\bibfnamefont {M.}~\bibnamefont
  {Piarulli}}, \bibinfo {author} {\bibfnamefont {S.}~\bibnamefont {Pastore}},
  \bibinfo {author} {\bibfnamefont {R.~B.}\ \bibnamefont {Wiringa}}, \bibinfo
  {author} {\bibfnamefont {S.}~\bibnamefont {Brusilow}},\ and\ \bibinfo
  {author} {\bibfnamefont {R.}~\bibnamefont {Lim}},\ }\bibfield  {title}
  {\bibinfo {title} {{Densities and momentum distributions in $A \le 12$ nuclei
  from chiral effective field theory interactions}},\ }\href
  {https://doi.org/10.1103/PhysRevC.107.014314} {\bibfield  {journal} {\bibinfo
   {journal} {Phys. Rev. C}\ }\textbf {\bibinfo {volume} {107}},\ \bibinfo
  {pages} {014314} (\bibinfo {year} {2023})}\BibitemShut {NoStop}%
\bibitem [{\citenamefont {Lonardoni}\ \emph {et~al.}(2017)\citenamefont
  {Lonardoni}, \citenamefont {Lovato}, \citenamefont {Pieper},\ and\
  \citenamefont {Wiringa}}]{lonardoni2017variational}%
  \BibitemOpen
  \bibfield  {author} {\bibinfo {author} {\bibfnamefont {D.}~\bibnamefont
  {Lonardoni}}, \bibinfo {author} {\bibfnamefont {A.}~\bibnamefont {Lovato}},
  \bibinfo {author} {\bibfnamefont {S.~C.}\ \bibnamefont {Pieper}},\ and\
  \bibinfo {author} {\bibfnamefont {R.~B.}\ \bibnamefont {Wiringa}},\
  }\bibfield  {title} {\bibinfo {title} {{Variational calculation of the ground
  state of closed-shell nuclei up to $A = 40$}},\ }\href
  {https://doi.org/10.1103/PhysRevC.96.024326} {\bibfield  {journal} {\bibinfo
  {journal} {Phys. Rev. C}\ }\textbf {\bibinfo {volume} {96}},\ \bibinfo
  {pages} {024326} (\bibinfo {year} {2017})}\BibitemShut {NoStop}%
\bibitem [{\citenamefont {Friman-Gayer}\ \emph {et~al.}(2021)\citenamefont
  {Friman-Gayer}, \citenamefont {Romig}, \citenamefont {H{\"u}ther},
  \citenamefont {Albe}, \citenamefont {Bacca}, \citenamefont {Beck},
  \citenamefont {Berger}, \citenamefont {Birkhan}, \citenamefont {Hebeler}
  \emph {et~al.}}]{friman2021role}%
  \BibitemOpen
  \bibfield  {author} {\bibinfo {author} {\bibfnamefont {U.}~\bibnamefont
  {Friman-Gayer}}, \bibinfo {author} {\bibfnamefont {C.}~\bibnamefont {Romig}},
  \bibinfo {author} {\bibfnamefont {T.}~\bibnamefont {H{\"u}ther}}, \bibinfo
  {author} {\bibfnamefont {K.}~\bibnamefont {Albe}}, \bibinfo {author}
  {\bibfnamefont {S.}~\bibnamefont {Bacca}}, \bibinfo {author} {\bibfnamefont
  {T.}~\bibnamefont {Beck}}, \bibinfo {author} {\bibfnamefont {M.}~\bibnamefont
  {Berger}}, \bibinfo {author} {\bibfnamefont {J.}~\bibnamefont {Birkhan}},
  \bibinfo {author} {\bibfnamefont {K.}~\bibnamefont {Hebeler}}, \emph
  {et~al.},\ }\bibfield  {title} {\bibinfo {title} {{Role of Chiral Two-Body
  Currents in $^{6}\mathrm{Li}$ Magnetic Properties in Light of a New Precision
  Measurement with the Relative Self-Absorption Technique}},\ }\href
  {https://doi.org/10.1103/PhysRevLett.126.102501} {\bibfield  {journal}
  {\bibinfo  {journal} {Phys. Rev. Lett.}\ }\textbf {\bibinfo {volume} {126}},\
  \bibinfo {pages} {102501} (\bibinfo {year} {2021})}\BibitemShut {NoStop}%
\bibitem [{\citenamefont {Jurgenson}\ \emph {et~al.}(2009)\citenamefont
  {Jurgenson}, \citenamefont {Navr\'{a}til},\ and\ \citenamefont
  {Furnstahl}}]{Jurgenson2009SRGManyBodyForces}%
  \BibitemOpen
  \bibfield  {author} {\bibinfo {author} {\bibfnamefont {E.~D.}\ \bibnamefont
  {Jurgenson}}, \bibinfo {author} {\bibfnamefont {P.}~\bibnamefont
  {Navr\'{a}til}},\ and\ \bibinfo {author} {\bibfnamefont {R.~J.}\ \bibnamefont
  {Furnstahl}},\ }\bibfield  {title} {\bibinfo {title} {{Evolution of Nuclear
  Many-Body Forces with the Similarity Renormalization Group}},\ }\href
  {https://doi.org/10.1103/PhysRevLett.103.082501} {\bibfield  {journal}
  {\bibinfo  {journal} {Phys. Rev. Lett.}\ }\textbf {\bibinfo {volume} {103}},\
  \bibinfo {pages} {082501} (\bibinfo {year} {2009})}\BibitemShut {NoStop}%
\bibitem [{\citenamefont {Jurgenson}\ \emph {et~al.}(2011)\citenamefont
  {Jurgenson}, \citenamefont {Navr\'{a}til},\ and\ \citenamefont
  {Furnstahl}}]{jurgenson2011evolving}%
  \BibitemOpen
  \bibfield  {author} {\bibinfo {author} {\bibfnamefont {E.~D.}\ \bibnamefont
  {Jurgenson}}, \bibinfo {author} {\bibfnamefont {P.}~\bibnamefont
  {Navr\'{a}til}},\ and\ \bibinfo {author} {\bibfnamefont {R.~J.}\ \bibnamefont
  {Furnstahl}},\ }\bibfield  {title} {\bibinfo {title} {{Evolving nuclear
  many-body forces with the similarity renormalization group}},\ }\href
  {https://doi.org/10.1103/PhysRevC.83.034301} {\bibfield  {journal} {\bibinfo
  {journal} {Phys. Rev. C}\ }\textbf {\bibinfo {volume} {83}},\ \bibinfo
  {pages} {034301} (\bibinfo {year} {2011})}\BibitemShut {NoStop}%
\bibitem [{\citenamefont {Roth}\ \emph {et~al.}(2014)\citenamefont {Roth},
  \citenamefont {Calci}, \citenamefont {Langhammer},\ and\ \citenamefont
  {Binder}}]{roth2014evolved}%
  \BibitemOpen
  \bibfield  {author} {\bibinfo {author} {\bibfnamefont {R.}~\bibnamefont
  {Roth}}, \bibinfo {author} {\bibfnamefont {A.}~\bibnamefont {Calci}},
  \bibinfo {author} {\bibfnamefont {J.}~\bibnamefont {Langhammer}},\ and\
  \bibinfo {author} {\bibfnamefont {S.}~\bibnamefont {Binder}},\ }\bibfield
  {title} {\bibinfo {title} {{Evolved chiral $NN + 3N$ Hamiltonians for ab
  initio nuclear structure calculations}},\ }\href
  {https://doi.org/10.1103/PhysRevC.90.024325} {\bibfield  {journal} {\bibinfo
  {journal} {Phys. Rev. C}\ }\textbf {\bibinfo {volume} {90}},\ \bibinfo
  {pages} {024325} (\bibinfo {year} {2014})}\BibitemShut {NoStop}%
\bibitem [{\citenamefont {Hebeler}(2021)}]{hebeler2021three}%
  \BibitemOpen
  \bibfield  {author} {\bibinfo {author} {\bibfnamefont {K.}~\bibnamefont
  {Hebeler}},\ }\bibfield  {title} {\bibinfo {title} {{Three-nucleon forces:
  Implementation and applications to atomic nuclei and dense matter}},\ }\href
  {https://doi.org/10.1016/j.physrep.2020.08.009} {\bibfield  {journal}
  {\bibinfo  {journal} {Phys. Rep.}\ }\textbf {\bibinfo {volume} {890}},\
  \bibinfo {pages} {1} (\bibinfo {year} {2021})}\BibitemShut {NoStop}%
\bibitem [{\citenamefont {Miyagi}\ \emph {et~al.}(2022)\citenamefont {Miyagi},
  \citenamefont {Stroberg}, \citenamefont {Navr{\'a}til}, \citenamefont
  {Hebeler},\ and\ \citenamefont {Holt}}]{miyagi2022converged}%
  \BibitemOpen
  \bibfield  {author} {\bibinfo {author} {\bibfnamefont {T.}~\bibnamefont
  {Miyagi}}, \bibinfo {author} {\bibfnamefont {S.~R.}\ \bibnamefont
  {Stroberg}}, \bibinfo {author} {\bibfnamefont {P.}~\bibnamefont
  {Navr{\'a}til}}, \bibinfo {author} {\bibfnamefont {K.}~\bibnamefont
  {Hebeler}},\ and\ \bibinfo {author} {\bibfnamefont {J.~D.}\ \bibnamefont
  {Holt}},\ }\bibfield  {title} {\bibinfo {title} {{Converged ab initio
  calculations of heavy nuclei}},\ }\href
  {https://doi.org/10.1103/PhysRevC.105.014302} {\bibfield  {journal} {\bibinfo
   {journal} {Phys. Rev. C}\ }\textbf {\bibinfo {volume} {105}},\ \bibinfo
  {pages} {014302} (\bibinfo {year} {2022})}\BibitemShut {NoStop}%
\bibitem [{\citenamefont {Schuster}\ \emph {et~al.}(2014)\citenamefont
  {Schuster}, \citenamefont {Quaglioni}, \citenamefont {Johnson}, \citenamefont
  {Jurgenson},\ and\ \citenamefont {Navr{\'a}til}}]{schuster2014operator}%
  \BibitemOpen
  \bibfield  {author} {\bibinfo {author} {\bibfnamefont {M.~D.}\ \bibnamefont
  {Schuster}}, \bibinfo {author} {\bibfnamefont {S.}~\bibnamefont {Quaglioni}},
  \bibinfo {author} {\bibfnamefont {C.~W.}\ \bibnamefont {Johnson}}, \bibinfo
  {author} {\bibfnamefont {E.~D.}\ \bibnamefont {Jurgenson}},\ and\ \bibinfo
  {author} {\bibfnamefont {P.}~\bibnamefont {Navr{\'a}til}},\ }\bibfield
  {title} {\bibinfo {title} {{Operator evolution for ab initio theory of light
  nuclei}},\ }\href {https://doi.org/10.1103/PhysRevC.90.011301} {\bibfield
  {journal} {\bibinfo  {journal} {Phys. Rev. C}\ }\textbf {\bibinfo {volume}
  {90}},\ \bibinfo {pages} {011301} (\bibinfo {year} {2014})}\BibitemShut
  {NoStop}%
\bibitem [{\citenamefont {Patsyuk}\ \emph {et~al.}(2021)\citenamefont
  {Patsyuk}, \citenamefont {Kahlbow}, \citenamefont {Laskaris}, \citenamefont
  {Duer}, \citenamefont {Lenivenko}, \citenamefont {Segarra}, \citenamefont
  {Atovullaev}, \citenamefont {Johansson}, \citenamefont {Aumann},
  \citenamefont {Corsi} \emph {et~al.}}]{patsyuk2021unperturbed}%
  \BibitemOpen
  \bibfield  {author} {\bibinfo {author} {\bibfnamefont {M.}~\bibnamefont
  {Patsyuk}}, \bibinfo {author} {\bibfnamefont {J.}~\bibnamefont {Kahlbow}},
  \bibinfo {author} {\bibfnamefont {G.}~\bibnamefont {Laskaris}}, \bibinfo
  {author} {\bibfnamefont {M.}~\bibnamefont {Duer}}, \bibinfo {author}
  {\bibfnamefont {V.}~\bibnamefont {Lenivenko}}, \bibinfo {author}
  {\bibfnamefont {E.}~\bibnamefont {Segarra}}, \bibinfo {author} {\bibfnamefont
  {T.}~\bibnamefont {Atovullaev}}, \bibinfo {author} {\bibfnamefont
  {G.}~\bibnamefont {Johansson}}, \bibinfo {author} {\bibfnamefont
  {T.}~\bibnamefont {Aumann}}, \bibinfo {author} {\bibfnamefont
  {A.}~\bibnamefont {Corsi}}, \emph {et~al.},\ }\bibfield  {title} {\bibinfo
  {title} {{Unperturbed inverse kinematics nucleon knockout measurements with a
  carbon beam}},\ }\href {https://doi.org/10.1038/s41567-021-01193-4}
  {\bibfield  {journal} {\bibinfo  {journal} {Nat. Phys.}\ }\textbf {\bibinfo
  {volume} {17}},\ \bibinfo {pages} {693} (\bibinfo {year} {2021})}\BibitemShut
  {NoStop}%
\bibitem [{\citenamefont {Ye}\ \emph {et~al.}(2024)\citenamefont {Ye},
  \citenamefont {Zhang}, \citenamefont {Zhang},\ and\ \citenamefont
  {Zhao}}]{ye2024new}%
  \BibitemOpen
  \bibfield  {author} {\bibinfo {author} {\bibfnamefont {Z.}~\bibnamefont
  {Ye}}, \bibinfo {author} {\bibfnamefont {H.}~\bibnamefont {Zhang}}, \bibinfo
  {author} {\bibfnamefont {Y.}~\bibnamefont {Zhang}},\ and\ \bibinfo {author}
  {\bibfnamefont {H.}~\bibnamefont {Zhao}},\ }\bibfield  {title} {\bibinfo
  {title} {{New Chinese facilities for short-range correlation physics}},\
  }\href {https://doi.org/10.1140/epja/s10050-024-01343-1} {\bibfield
  {journal} {\bibinfo  {journal} {Eur. Phys. J. A}\ }\textbf {\bibinfo {volume}
  {60}},\ \bibinfo {pages} {126} (\bibinfo {year} {2024})}\BibitemShut
  {NoStop}%
\bibitem [{\citenamefont {Tsukiyama}\ \emph {et~al.}(2011)\citenamefont
  {Tsukiyama}, \citenamefont {Bogner},\ and\ \citenamefont
  {Schwenk}}]{tsukiyama2011medium}%
  \BibitemOpen
  \bibfield  {author} {\bibinfo {author} {\bibfnamefont {K.}~\bibnamefont
  {Tsukiyama}}, \bibinfo {author} {\bibfnamefont {S.~K.}\ \bibnamefont
  {Bogner}},\ and\ \bibinfo {author} {\bibfnamefont {A.}~\bibnamefont
  {Schwenk}},\ }\bibfield  {title} {\bibinfo {title} {{In-medium similarity
  renormalization group for nuclei}},\ }\href
  {https://doi.org/10.1103/PhysRevLett.106.222502} {\bibfield  {journal}
  {\bibinfo  {journal} {Phys. Rev. Lett.}\ }\textbf {\bibinfo {volume} {106}},\
  \bibinfo {pages} {222502} (\bibinfo {year} {2011})}\BibitemShut {NoStop}%
\bibitem [{\citenamefont {Hergert}\ \emph {et~al.}(2016)\citenamefont
  {Hergert}, \citenamefont {Bogner}, \citenamefont {Morris}, \citenamefont
  {Schwenk},\ and\ \citenamefont {Tsukiyama}}]{hergert2016medium}%
  \BibitemOpen
  \bibfield  {author} {\bibinfo {author} {\bibfnamefont {H.}~\bibnamefont
  {Hergert}}, \bibinfo {author} {\bibfnamefont {S.~K.}\ \bibnamefont {Bogner}},
  \bibinfo {author} {\bibfnamefont {T.~D.}\ \bibnamefont {Morris}}, \bibinfo
  {author} {\bibfnamefont {A.}~\bibnamefont {Schwenk}},\ and\ \bibinfo {author}
  {\bibfnamefont {K.}~\bibnamefont {Tsukiyama}},\ }\bibfield  {title} {\bibinfo
  {title} {{The in-medium similarity renormalization group: A novel $\textit{ab
  initio}$ method for nuclei}},\ }\href
  {https://doi.org/10.1016/j.physrep.2015.12.007} {\bibfield  {journal}
  {\bibinfo  {journal} {Phys. Rep.}\ }\textbf {\bibinfo {volume} {621}},\
  \bibinfo {pages} {165} (\bibinfo {year} {2016})}\BibitemShut {NoStop}%
\bibitem [{\citenamefont {Tsukiyama}\ \emph {et~al.}(2012)\citenamefont
  {Tsukiyama}, \citenamefont {Bogner},\ and\ \citenamefont
  {Schwenk}}]{tsukiyama2012medium}%
  \BibitemOpen
  \bibfield  {author} {\bibinfo {author} {\bibfnamefont {K.}~\bibnamefont
  {Tsukiyama}}, \bibinfo {author} {\bibfnamefont {S.~K.}\ \bibnamefont
  {Bogner}},\ and\ \bibinfo {author} {\bibfnamefont {A.}~\bibnamefont
  {Schwenk}},\ }\bibfield  {title} {\bibinfo {title} {{In-medium similarity
  renormalization group for open-shell nuclei}},\ }\href
  {https://doi.org/10.1103/PhysRevC.85.061304} {\bibfield  {journal} {\bibinfo
  {journal} {Phys. Rev. C}\ }\textbf {\bibinfo {volume} {85}},\ \bibinfo
  {pages} {061304} (\bibinfo {year} {2012})}\BibitemShut {NoStop}%
\bibitem [{\citenamefont {Cruz-Torres}\ \emph {et~al.}(2018)\citenamefont
  {Cruz-Torres}, \citenamefont {Schmidt}, \citenamefont {Miller}, \citenamefont
  {Weinstein}, \citenamefont {Barnea}, \citenamefont {Weiss}, \citenamefont
  {Piasetzky},\ and\ \citenamefont {Hen}}]{cruz2018short}%
  \BibitemOpen
  \bibfield  {author} {\bibinfo {author} {\bibfnamefont {R.}~\bibnamefont
  {Cruz-Torres}}, \bibinfo {author} {\bibfnamefont {A.}~\bibnamefont
  {Schmidt}}, \bibinfo {author} {\bibfnamefont {G.~A.}\ \bibnamefont {Miller}},
  \bibinfo {author} {\bibfnamefont {L.~B.}\ \bibnamefont {Weinstein}}, \bibinfo
  {author} {\bibfnamefont {N.}~\bibnamefont {Barnea}}, \bibinfo {author}
  {\bibfnamefont {R.}~\bibnamefont {Weiss}}, \bibinfo {author} {\bibfnamefont
  {E.}~\bibnamefont {Piasetzky}},\ and\ \bibinfo {author} {\bibfnamefont
  {O.}~\bibnamefont {Hen}},\ }\bibfield  {title} {\bibinfo {title} {{Short
  range correlations and the isospin dependence of nuclear correlation
  functions}},\ }\href {https://doi.org/10.1016/j.physletb.2018.07.069}
  {\bibfield  {journal} {\bibinfo  {journal} {Phys. Lett. B}\ }\textbf
  {\bibinfo {volume} {785}},\ \bibinfo {pages} {304} (\bibinfo {year}
  {2018})}\BibitemShut {NoStop}%
\bibitem [{\citenamefont {Xu}\ \emph {et~al.}(2024)\citenamefont {Xu},
  \citenamefont {Fan}, \citenamefont {Yuan}, \citenamefont {Hu}, \citenamefont
  {Li}, \citenamefont {Wang},\ and\ \citenamefont {Xu}}]{xu2024progress}%
  \BibitemOpen
  \bibfield  {author} {\bibinfo {author} {\bibfnamefont {X.-Y.}\ \bibnamefont
  {Xu}}, \bibinfo {author} {\bibfnamefont {S.-Q.}\ \bibnamefont {Fan}},
  \bibinfo {author} {\bibfnamefont {Q.}~\bibnamefont {Yuan}}, \bibinfo {author}
  {\bibfnamefont {B.-S.}\ \bibnamefont {Hu}}, \bibinfo {author} {\bibfnamefont
  {J.-G.}\ \bibnamefont {Li}}, \bibinfo {author} {\bibfnamefont {S.-M.}\
  \bibnamefont {Wang}},\ and\ \bibinfo {author} {\bibfnamefont {F.-R.}\
  \bibnamefont {Xu}},\ }\bibfield  {title} {\bibinfo {title} {{Progress in ab
  initio in-medium similarity renormalization group and coupled-channel method
  with coupling to the continuum}},\ }\href
  {https://doi.org/10.1007/s41365-024-01585-0} {\bibfield  {journal} {\bibinfo
  {journal} {Nucl. Sci. Tech.}\ }\textbf {\bibinfo {volume} {35}},\ \bibinfo
  {pages} {215} (\bibinfo {year} {2024})}\BibitemShut {NoStop}%
\bibitem [{\citenamefont {Roth}\ \emph {et~al.}(2011)\citenamefont {Roth},
  \citenamefont {Langhammer}, \citenamefont {Calci}, \citenamefont {Binder},\
  and\ \citenamefont {Navr{\'a}til}}]{roth2011similarity}%
  \BibitemOpen
  \bibfield  {author} {\bibinfo {author} {\bibfnamefont {R.}~\bibnamefont
  {Roth}}, \bibinfo {author} {\bibfnamefont {J.}~\bibnamefont {Langhammer}},
  \bibinfo {author} {\bibfnamefont {A.}~\bibnamefont {Calci}}, \bibinfo
  {author} {\bibfnamefont {S.}~\bibnamefont {Binder}},\ and\ \bibinfo {author}
  {\bibfnamefont {P.}~\bibnamefont {Navr{\'a}til}},\ }\bibfield  {title}
  {\bibinfo {title} {{Similarity-Transformed Chiral $NN + 3N$ Interactions for
  the $\textit{Ab Initio}$ Description of $^{12}\mathrm{C}$ and
  $^{16}\mathrm{O}$}},\ }\href {https://doi.org/10.1103/PhysRevLett.107.072501}
  {\bibfield  {journal} {\bibinfo  {journal} {Phys. Rev. Lett.}\ }\textbf
  {\bibinfo {volume} {107}},\ \bibinfo {pages} {072501} (\bibinfo {year}
  {2011})}\BibitemShut {NoStop}%
\bibitem [{\citenamefont {Roth}\ \emph {et~al.}(2012)\citenamefont {Roth},
  \citenamefont {Binder}, \citenamefont {Vobig}, \citenamefont {Calci},
  \citenamefont {Langhammer},\ and\ \citenamefont
  {Navr{\'a}til}}]{roth2012medium}%
  \BibitemOpen
  \bibfield  {author} {\bibinfo {author} {\bibfnamefont {R.}~\bibnamefont
  {Roth}}, \bibinfo {author} {\bibfnamefont {S.}~\bibnamefont {Binder}},
  \bibinfo {author} {\bibfnamefont {K.}~\bibnamefont {Vobig}}, \bibinfo
  {author} {\bibfnamefont {A.}~\bibnamefont {Calci}}, \bibinfo {author}
  {\bibfnamefont {J.}~\bibnamefont {Langhammer}},\ and\ \bibinfo {author}
  {\bibfnamefont {P.}~\bibnamefont {Navr{\'a}til}},\ }\bibfield  {title}
  {\bibinfo {title} {{Medium-Mass Nuclei with Normal-Ordered Chiral $NN + 3N$
  Interactions}},\ }\href {https://doi.org/10.1103/PhysRevLett.109.052501}
  {\bibfield  {journal} {\bibinfo  {journal} {Phys. Rev. Lett.}\ }\textbf
  {\bibinfo {volume} {109}},\ \bibinfo {pages} {052501} (\bibinfo {year}
  {2012})}\BibitemShut {NoStop}%
\bibitem [{\citenamefont {White}(2002)}]{white2002numerical}%
  \BibitemOpen
  \bibfield  {author} {\bibinfo {author} {\bibfnamefont {S.~R.}\ \bibnamefont
  {White}},\ }\bibfield  {title} {\bibinfo {title} {{Numerical canonical
  transformation approach to quantum many-body problems}},\ }\href
  {https://doi.org/10.1063/1.1508370} {\bibfield  {journal} {\bibinfo
  {journal} {J. Chem. Phys.}\ }\textbf {\bibinfo {volume} {117}},\ \bibinfo
  {pages} {7472–7482} (\bibinfo {year} {2002})}\BibitemShut {NoStop}%
\bibitem [{\citenamefont {Entem}\ \emph {et~al.}(2017)\citenamefont {Entem},
  \citenamefont {Machleidt},\ and\ \citenamefont {Nosyk}}]{entem2017high}%
  \BibitemOpen
  \bibfield  {author} {\bibinfo {author} {\bibfnamefont {D.~R.}\ \bibnamefont
  {Entem}}, \bibinfo {author} {\bibfnamefont {R.}~\bibnamefont {Machleidt}},\
  and\ \bibinfo {author} {\bibfnamefont {Y.}~\bibnamefont {Nosyk}},\ }\bibfield
   {title} {\bibinfo {title} {{High-quality two-nucleon potentials up to fifth
  order of the chiral expansion}},\ }\href
  {https://doi.org/10.1103/PhysRevC.96.024004} {\bibfield  {journal} {\bibinfo
  {journal} {Phys. Rev. C}\ }\textbf {\bibinfo {volume} {96}},\ \bibinfo
  {pages} {024004} (\bibinfo {year} {2017})}\BibitemShut {NoStop}%
\bibitem [{\citenamefont {Gysbers}\ \emph {et~al.}(2019)\citenamefont
  {Gysbers}, \citenamefont {Hagen}, \citenamefont {Holt}, \citenamefont
  {Jansen}, \citenamefont {Morris}, \citenamefont {Navr{\'a}til}, \citenamefont
  {Papenbrock}, \citenamefont {Quaglioni}, \citenamefont {Schwenk},
  \citenamefont {Stroberg} \emph {et~al.}}]{gysbers2019discrepancy}%
  \BibitemOpen
  \bibfield  {author} {\bibinfo {author} {\bibfnamefont {P.}~\bibnamefont
  {Gysbers}}, \bibinfo {author} {\bibfnamefont {G.}~\bibnamefont {Hagen}},
  \bibinfo {author} {\bibfnamefont {J.~D.}\ \bibnamefont {Holt}}, \bibinfo
  {author} {\bibfnamefont {G.~R.}\ \bibnamefont {Jansen}}, \bibinfo {author}
  {\bibfnamefont {T.~D.}\ \bibnamefont {Morris}}, \bibinfo {author}
  {\bibfnamefont {P.}~\bibnamefont {Navr{\'a}til}}, \bibinfo {author}
  {\bibfnamefont {T.}~\bibnamefont {Papenbrock}}, \bibinfo {author}
  {\bibfnamefont {S.}~\bibnamefont {Quaglioni}}, \bibinfo {author}
  {\bibfnamefont {A.}~\bibnamefont {Schwenk}}, \bibinfo {author} {\bibfnamefont
  {S.~R.}\ \bibnamefont {Stroberg}}, \emph {et~al.},\ }\bibfield  {title}
  {\bibinfo {title} {{Discrepancy between experimental and theoretical
  $\beta$-decay rates resolved from first principles}},\ }\href
  {https://doi.org/10.1038/s41567-019-0450-7} {\bibfield  {journal} {\bibinfo
  {journal} {Nat. Phys.}\ }\textbf {\bibinfo {volume} {15}},\ \bibinfo {pages}
  {428} (\bibinfo {year} {2019})}\BibitemShut {NoStop}%
\bibitem [{\citenamefont {Miyagi}(2023)}]{miyagi2023nuhamil}%
  \BibitemOpen
  \bibfield  {author} {\bibinfo {author} {\bibfnamefont {T.}~\bibnamefont
  {Miyagi}},\ }\bibfield  {title} {\bibinfo {title} {{NuHamil: A numerical code
  to generate nuclear two-and three-body matrix elements from chiral effective
  field theory}},\ }\href {https://doi.org/10.1140/epja/s10050-023-01039-y}
  {\bibfield  {journal} {\bibinfo  {journal} {Eur. Phys. J. A}\ }\textbf
  {\bibinfo {volume} {59}},\ \bibinfo {pages} {150} (\bibinfo {year}
  {2023})}\BibitemShut {NoStop}%
\bibitem [{\citenamefont {Stroberg}()}]{StrobergIMSRG2021}%
  \BibitemOpen
  \bibfield  {author} {\bibinfo {author} {\bibfnamefont {S.~R.}\ \bibnamefont
  {Stroberg}},\ }\href {https://github.com/ragnarstroberg/imsrg} {}\bibinfo
  {howpublished} {\url{https://github.com/ragnarstroberg/imsrg}}\BibitemShut
  {NoStop}%
\bibitem [{\citenamefont {Shimizu}\ \emph {et~al.}(2019)\citenamefont
  {Shimizu}, \citenamefont {Mizusaki}, \citenamefont {Utsuno},\ and\
  \citenamefont {Tsunoda}}]{shimizu2019thick}%
  \BibitemOpen
  \bibfield  {author} {\bibinfo {author} {\bibfnamefont {N.}~\bibnamefont
  {Shimizu}}, \bibinfo {author} {\bibfnamefont {T.}~\bibnamefont {Mizusaki}},
  \bibinfo {author} {\bibfnamefont {Y.}~\bibnamefont {Utsuno}},\ and\ \bibinfo
  {author} {\bibfnamefont {Y.}~\bibnamefont {Tsunoda}},\ }\bibfield  {title}
  {\bibinfo {title} {{Thick-restart block Lanczos method for large-scale
  shell-model calculations}},\ }\href
  {https://doi.org/10.1016/j.cpc.2019.06.011} {\bibfield  {journal} {\bibinfo
  {journal} {Comput. Phys. Commun.}\ }\textbf {\bibinfo {volume} {244}},\
  \bibinfo {pages} {372} (\bibinfo {year} {2019})}\BibitemShut {NoStop}%
\bibitem [{\citenamefont {Barrett}\ \emph {et~al.}(2013)\citenamefont
  {Barrett}, \citenamefont {Navr{\'a}til},\ and\ \citenamefont
  {Vary}}]{barrett2013ab}%
  \BibitemOpen
  \bibfield  {author} {\bibinfo {author} {\bibfnamefont {B.~R.}\ \bibnamefont
  {Barrett}}, \bibinfo {author} {\bibfnamefont {P.}~\bibnamefont
  {Navr{\'a}til}},\ and\ \bibinfo {author} {\bibfnamefont {J.~P.}\ \bibnamefont
  {Vary}},\ }\bibfield  {title} {\bibinfo {title} {{Ab initio no core shell
  model}},\ }\href {https://doi.org/10.1016/j.ppnp.2012.10.003} {\bibfield
  {journal} {\bibinfo  {journal} {Prog. Part. Nucl. Phys.}\ }\textbf {\bibinfo
  {volume} {69}},\ \bibinfo {pages} {131} (\bibinfo {year} {2013})}\BibitemShut
  {NoStop}%
\bibitem [{\citenamefont {Feldmeier}\ \emph {et~al.}(2011)\citenamefont
  {Feldmeier}, \citenamefont {Horiuchi}, \citenamefont {Neff},\ and\
  \citenamefont {Suzuki}}]{feldmeier2011universality}%
  \BibitemOpen
  \bibfield  {author} {\bibinfo {author} {\bibfnamefont {H.}~\bibnamefont
  {Feldmeier}}, \bibinfo {author} {\bibfnamefont {W.}~\bibnamefont {Horiuchi}},
  \bibinfo {author} {\bibfnamefont {T.}~\bibnamefont {Neff}},\ and\ \bibinfo
  {author} {\bibfnamefont {Y.}~\bibnamefont {Suzuki}},\ }\bibfield  {title}
  {\bibinfo {title} {{Universality of short-range nucleon-nucleon
  correlations}},\ }\href {https://doi.org/10.1103/PhysRevC.84.054003}
  {\bibfield  {journal} {\bibinfo  {journal} {Phys. Rev. C}\ }\textbf {\bibinfo
  {volume} {84}},\ \bibinfo {pages} {054003} (\bibinfo {year}
  {2011})}\BibitemShut {NoStop}%
\bibitem [{\citenamefont {Li}\ \emph {et~al.}(2025)\citenamefont {Li},
  \citenamefont {Santiesteban}, \citenamefont {Arrington}, \citenamefont
  {Cruz-Torres}, \citenamefont {Kurbany}, \citenamefont {Abrams}, \citenamefont
  {Alsalmi}, \citenamefont {Androic}, \citenamefont {Aniol}, \citenamefont
  {Averett} \emph {et~al.}}]{li2025inclusive}%
  \BibitemOpen
  \bibfield  {author} {\bibinfo {author} {\bibfnamefont {S.}~\bibnamefont
  {Li}}, \bibinfo {author} {\bibfnamefont {S.~N.}\ \bibnamefont
  {Santiesteban}}, \bibinfo {author} {\bibfnamefont {J.}~\bibnamefont
  {Arrington}}, \bibinfo {author} {\bibfnamefont {R.}~\bibnamefont
  {Cruz-Torres}}, \bibinfo {author} {\bibfnamefont {L.}~\bibnamefont
  {Kurbany}}, \bibinfo {author} {\bibfnamefont {D.}~\bibnamefont {Abrams}},
  \bibinfo {author} {\bibfnamefont {S.}~\bibnamefont {Alsalmi}}, \bibinfo
  {author} {\bibfnamefont {D.}~\bibnamefont {Androic}}, \bibinfo {author}
  {\bibfnamefont {K.}~\bibnamefont {Aniol}}, \bibinfo {author} {\bibfnamefont
  {T.}~\bibnamefont {Averett}}, \emph {et~al.},\ }\bibfield  {title} {\bibinfo
  {title} {{Inclusive studies of two-and three-nucleon short-range correlations
  in $^{3}\mathrm{H}$ and $^{3}\mathrm{He}$}},\ }\href
  {https://doi.org/10.1016/j.physletb.2025.139734} {\bibfield  {journal}
  {\bibinfo  {journal} {Phys. Lett. B}\ ,\ \bibinfo {pages} {139734}} (\bibinfo
  {year} {2025})}\BibitemShut {NoStop}%
\bibitem [{\citenamefont {Hen}\ \emph {et~al.}(2014)\citenamefont {Hen},
  \citenamefont {Sargsian}, \citenamefont {Weinstein}, \citenamefont
  {Piasetzky}, \citenamefont {Hakobyan}, \citenamefont {Higinbotham},
  \citenamefont {Braverman}, \citenamefont {Brooks}, \citenamefont {Gilad},
  \citenamefont {Adhikari} \emph {et~al.}}]{hen2014momentum}%
  \BibitemOpen
  \bibfield  {author} {\bibinfo {author} {\bibfnamefont {O.}~\bibnamefont
  {Hen}}, \bibinfo {author} {\bibfnamefont {M.}~\bibnamefont {Sargsian}},
  \bibinfo {author} {\bibfnamefont {L.~B.}\ \bibnamefont {Weinstein}}, \bibinfo
  {author} {\bibfnamefont {E.}~\bibnamefont {Piasetzky}}, \bibinfo {author}
  {\bibfnamefont {H.}~\bibnamefont {Hakobyan}}, \bibinfo {author}
  {\bibfnamefont {D.~W.}\ \bibnamefont {Higinbotham}}, \bibinfo {author}
  {\bibfnamefont {M.}~\bibnamefont {Braverman}}, \bibinfo {author}
  {\bibfnamefont {W.~K.}\ \bibnamefont {Brooks}}, \bibinfo {author}
  {\bibfnamefont {S.}~\bibnamefont {Gilad}}, \bibinfo {author} {\bibfnamefont
  {K.~P.}\ \bibnamefont {Adhikari}}, \emph {et~al.},\ }\bibfield  {title}
  {\bibinfo {title} {{Momentum sharing in imbalanced Fermi systems}},\ }\href
  {https://doi.org/10.1126/science.1256785} {\bibfield  {journal} {\bibinfo
  {journal} {Science}\ }\textbf {\bibinfo {volume} {346}},\ \bibinfo {pages}
  {614} (\bibinfo {year} {2014})}\BibitemShut {NoStop}%
\bibitem [{\citenamefont {Nguyen}\ \emph {et~al.}(2020)\citenamefont {Nguyen},
  \citenamefont {Ye}, \citenamefont {Aguilera}, \citenamefont {Ahmed},
  \citenamefont {Albataineh}, \citenamefont {Allada}, \citenamefont {Anderson},
  \citenamefont {Anez}, \citenamefont {Aniol}, \citenamefont {Annand} \emph
  {et~al.}}]{nguyen2020novel}%
  \BibitemOpen
  \bibfield  {author} {\bibinfo {author} {\bibfnamefont {D.}~\bibnamefont
  {Nguyen}}, \bibinfo {author} {\bibfnamefont {Z.}~\bibnamefont {Ye}}, \bibinfo
  {author} {\bibfnamefont {P.}~\bibnamefont {Aguilera}}, \bibinfo {author}
  {\bibfnamefont {Z.}~\bibnamefont {Ahmed}}, \bibinfo {author} {\bibfnamefont
  {H.}~\bibnamefont {Albataineh}}, \bibinfo {author} {\bibfnamefont
  {K.}~\bibnamefont {Allada}}, \bibinfo {author} {\bibfnamefont
  {B.}~\bibnamefont {Anderson}}, \bibinfo {author} {\bibfnamefont
  {D.}~\bibnamefont {Anez}}, \bibinfo {author} {\bibfnamefont {K.}~\bibnamefont
  {Aniol}}, \bibinfo {author} {\bibfnamefont {J.}~\bibnamefont {Annand}}, \emph
  {et~al.},\ }\bibfield  {title} {\bibinfo {title} {{Novel observation of
  isospin structure of short-range correlations in calcium isotopes}},\ }\href
  {https://doi.org/10.1103/PhysRevC.102.064004} {\bibfield  {journal} {\bibinfo
   {journal} {Phys. Rev. C}\ }\textbf {\bibinfo {volume} {102}},\ \bibinfo
  {pages} {064004} (\bibinfo {year} {2020})}\BibitemShut {NoStop}%
\end{thebibliography}%

\end{CJK*}
\end{document}